%% file: EWK-11-021_temp.tex
\pdfoutput=1

\documentclass[11pt,twoside,a4paper,cmspaper,final,collab]{cms-tdr}
\def\svnVersion{185181}\def\cmsCernNo{2013-001}\def\cmsCernDate{\today}\def\cmsMessage{Submitted to Physics Letters B}

\begin{document}\cmsNoteHeader{EWK-11-021}

\hyphenation{had-ron-i-za-tion}
\hyphenation{cal-or-i-me-ter}
\hyphenation{de-vices}

\RCS$Revision: 177216 $
\RCS$HeadURL: svn+ssh://svn.cern.ch/reps/tdr2/papers/EWK-11-021/trunk/EWK-11-021.tex $
\RCS$Id: EWK-11-021.tex 177216 2013-03-19 14:01:14Z theofil $

\newlength\figureheight
\newsavebox{\tempbox}
\newcommand{\setsubfloat}[3]
	{
	\includegraphics[width=#1]{#2}
	}
\newcommand{\setsubfloatR}[3]
	{
	\includegraphics[width=#1]{#2}
	}
\newcommand{\setsubfloatT}[3]
	{
	\includegraphics[width=#1]{#2}
	}
\newcommand{\setsubfloatTR}[3]
	{
	\includegraphics[width=#1]{#2}
	}

\newcommand{\plus}{\ensuremath{+}}
\newcommand{\vjets}{\ensuremath{\mathrm{V}}\text{ + jets}\xspace}
\newcommand{\met}{\ensuremath{\mspace{3mu}/\mspace{-12.0mu}E_{T}}}
\newcommand{\njets}{N_\ensuremath{\text{jets}}\xspace}
\newcommand{\lumifinal}{5.0\fbinv}
\newcommand{\mc}{Monte Carlo}
\newcommand{\MC}{\ensuremath{\textrm{MC}}\xspace}

\newcommand{\V}{\ensuremath{\mathrm{V}}\xspace}
\newcommand{\W}{\PW\xspace}
\renewcommand{\Z}{\cPZ\xspace}
\newcommand{\zjets}{\cPZ\text{ + jets}\xspace}
\newcommand{\Zmumu}{\cPZ\xspace\ensuremath{\rightarrow \mu\mu}\xspace}
\newcommand{\Znunu}{\cPZ\xspace\ensuremath{\rightarrow \nu\nu}\xspace}
\newcommand{\Zee}{\ensuremath{\cPZ \rightarrow \Pe \Pe}\xspace}
\newcommand{\Ztautau}{\ensuremath{\cPZ \rightarrow \Pgt \Pgt}\xspace}
\newcommand{\ptZ}{\ensuremath{p_\mathrm{T}^\cPZ}\xspace}
\renewcommand{\ttbar}{\ensuremath{\cPqt\cPaqt}\xspace}

\newcommand{\WW}{\ensuremath{\W\W}\xspace}
\newcommand{\WZ}{\ensuremath{\W\Z}\xspace}
\newcommand{\ZZ}{\ensuremath{\Z\Z}\xspace}

\newcommand{\fastJET}{\ensuremath{\textrm{\sc FastJet}}\xspace}

\newlength\cmsFigWidth
\ifthenelse{\boolean{cms@external}}{\setlength\cmsFigWidth{0.85\columnwidth}}{\setlength\cmsFigWidth{0.4\textwidth}}
\ifthenelse{\boolean{cms@external}}{\providecommand{\cmsLeft}{top}}{\providecommand{\cmsLeft}{left}}
\ifthenelse{\boolean{cms@external}}{\providecommand{\cmsRight}{bottom}}{\providecommand{\cmsRight}{right}}
\cmsNoteHeader{EWK-11-021}

\title{Event shapes and azimuthal correlations in \texorpdfstring{\cPZ\ + jets events in $\Pp\Pp$ collisions at $\sqrt{s}=7\TeV$}{Z+jets events in pp collisions at sqrt(s) = 7 TeV}}

\date{\today}
\abstract{
Measurements of event shapes and azimuthal correlations are presented for events where a \Z boson is produced in association with jets in proton-proton collisions.
The data collected with the CMS detector at the CERN LHC at $\sqrt{s}=7\TeV$ correspond to an integrated luminosity of 5.0\fbinv.
The analysis provides a test of predictions from perturbative QCD for a process that represents a substantial background to many physics channels.
Results are presented as a function of jet multiplicity, for inclusive \Z boson production and for \Z bosons with transverse momenta greater than 150\GeV, and
compared to predictions from Monte Carlo event generators that include leading-order multiparton matrix-element (with up to four hard partons in the final state)
and next-to-leading-order simulations of \Z~+~1-jet events.
The experimental results are corrected for detector effects, and can be compared directly with other QCD models.
}

\hypersetup{%
pdfauthor={CMS Collaboration},%
pdftitle={Event shapes and azimuthal correlations in Z + jets events in pp collisions at sqrt(s) =7 TeV},%
pdfsubject={CMS},%
pdfkeywords={CMS, physics, event shape, QCD}}

\maketitle

\section{Introduction}

A detailed study of the production of a \Z boson in association with jets in pp collisions
at the CERN Large Hadron Collider (LHC) is of great interest.
Measurements of this process can be confronted with the predictions of perturbative quantum chromodynamics (QCD) at the highest accessible energies and for a broad range of kinematic configurations.
Considerable theoretical progress has been made in this field,
such as developments in next-to-leading-order (NLO) calculations for up to four hard partons produced in
association with a \Z boson~\cite{Ita:2011wn}, NLO predictions
for \Z +~1-jet production that can be interfaced to parton shower (PS) approximations~\cite{powheg:2004,powheg:2007,Alioli:2010xd,Alioli:2010qp,MCatNLO},
and leading-order (LO) multiparton matrix-element (ME) event generators such as \ALPGEN~\cite{Mangano:2002ea}, \MADGRAPH~\cite{Alwall:2011uj}, and \SHERPA~\cite{sherpa}, with provision
for PS development.
In addition, \zjets production corresponds to a major background to many other processes at the LHC,
such as the production of top quarks, and it is important in searches for supersymmetric particles and Higgs boson physics.
An improved understanding of \zjets production over the largest possible regions of phase space can therefore provide a helpful tool for extracting small signals.

Previous studies of angular correlations between the \Z and the ``leading'' jet (the one with the largest \pt) and between the two jets of largest \pt have been reported at the Tevatron by the D0 Collaboration~\cite{d0vjets} and at the LHC by the ATLAS Collaboration using $36\pbinv$ of integrated luminosity~\cite{atlasvjets}.
In this Letter, the comparison of models with data for highly boosted \Z bosons with $\ptZ>150\GeV$ is of particular interest.
This region of phase space is critical in searches for new phenomena that are based on a large apparent imbalance in the total transverse momentum.
Such imbalance can be produced, \eg by the \Znunu standard model (SM) background.
The uncertainty of this background contribution is limited by the accuracy of current Monte Carlo (\MC) models, which can be improved through studies of leptonic $( \ell^+ \ell^-)$
decays of \Z bosons and their correlations with the associated jets.

In addition to azimuthal distributions, we provide the first measurements of variables that categorize the topological structure of \zjets events.
Multijet production at \EE and $\Pe\Pp$ colliders was used in the past to tune parton showers and fragmentation functions in \MC event generators,
as well as to measure the values of the strong coupling constant \cite{H1evshp,Zeusevshp,DelphiTuningMCevshp,dissertorievshp,AlephQCDoverview}.
A set of event-shape variables suitable for hadron colliders has been proposed in Ref.~\cite{evshpSalam}, which provides resummed perturbative predictions at next-to-leading-log (NLL) for these
variables. A measurement of event shapes in multijet events was reported recently by the Compact Muon Solenoid (CMS) Collaboration~\cite{evshpPaper}.

This Letter extends measurements of angular correlations and event shapes in \zjets events by probing the features of final states containing $\Z\rightarrow\ell^+\ell^-$ decays,
where $\ell=\mu$ or $\Pe$.
Such final states, often referred to as Drell--Yan (DY), include contributions from $\Pgg^*$ and $\Z/\Pgg^*$ interference terms arising from the irreducible background of virtual photons $(\Pgg^*)$ from
$\cPq\cPaq \rightarrow \Pgg^* \rightarrow \ell^+\ell^-$ processes.
The data were collected with the CMS detector at a center-of-mass energy of $7\TeV$, and correspond to an integrated luminosity of \lumifinal.
The observed angular distributions and event shapes in \zjets production are compared with predictions from several \MC generators, and comprise the first study of this kind
to be reported at the LHC.

\section{CMS detector}

The origin of the CMS coordinate system is chosen at the center of the detector, with the $z$ axis pointing along the direction of the counterclockwise proton beam.
The azimuthal angle is denoted as $\phi$, the polar angle as $\theta$, and the pseudorapidity is defined as $\eta=-\ln\left[\tan\left(\theta/2\right)\right]$.
The central feature of the CMS detector is a superconducting solenoid of 6\unit{m} internal diameter that produces an axial magnetic field of 3.8\unit{T}.
A silicon pixel and strip tracker, a lead tungstate crystal electromagnetic calorimeter (ECAL), and a brass/plastic-scintillator hadronic calorimeter (HCAL) are positioned within the field volume.
Iron and quartz-fiber hadronic calorimeters are located outside the magnetic field volume, within each endcap region of the CMS detector, at $3<|\eta|<5$.
Muons are measured using gas-ionization detectors embedded in the flux-return yoke outside of the solenoid.
A detailed description of the CMS detector can be found in Ref.~\cite{Chatrchyan:2008zzk}.

\section{Monte Carlo simulation}
All production processes of concern, namely the \zjets signal and backgrounds corresponding to top-antitop quark pairs (\ttbar), dibosons (\WZ, \ZZ, \WW), single top, and \W + jets events
are generated with \MADGRAPH (version 5.1.1.0)~\cite{Alwall:2011uj}, which provides up to four-parton final states and is interfaced to \PYTHIA (version 6.4.24)~\cite{pythia6}
using the Z2 tune~\cite{Chatrchyan:2011id} to implement showering and hadronization of the partons.
The CTEQ6L1~\cite{Pumplin:2002vw} parton distribution functions (PDF) are chosen for these calculations.
Alternative models for signal include (i) \SHERPA (version 1.3.1)~\cite{sherpa} (with up to four-parton final states)
using the CTEQ6m PDF~\cite{Pumplin:2002vw} and the default tune, (ii) \POWHEG \cite{powheg:2004,powheg:2007,Alioli:2010xd,Alioli:2010qp} for generating \Z + 1-jet events at NLO using
the CT10 PDF~\cite{Lai:2010vv} and interfaced to \PYTHIA (version 6.4.24) with the Z2 tune for parton showering and hadronization,
and (iii) stand-alone \PYTHIA (version 6.4.24) with the Z2 tune.
The cross sections for electroweak \Z and \W boson production are normalized to match the next-to-next-to-leading-order (NNLO) prediction obtained
with FEWZ~\cite{nnlo:Z} and the CTEQ6m PDF.
The \ttbar cross section is normalized to the next-to-next-to-leading-log (NNLL) calculation from Ref.~\cite{ttbarxsec}.
NLO precision obtained from \MCFM~\cite{Campbell:2010ff} with CTEQ6m PDF is used for the cross section for diboson (\WZ, \ZZ, \WW) and single top processes.

The detector response is simulated using a detailed description of the CMS detector based on the \GEANTfour package~\cite{geant4}, and the \MC simulated events are reconstructed
following the same procedures as used for data.
During the data taking, an average of ten minimum-bias interactions occurred in each bunch crossing (pileup).
The prevailing beam conditions are taken into account by reweighting the \MC simulation to match the spectrum of pileup interactions observed in data.

\section{Event selection and reconstruction}
\label{eventreco}

Event selection starts by requiring two high-\pt leptons at the trigger level.
For muons, this corresponds to an online \pt threshold of $13\GeV$ ($17\GeV$ during periods of higher instantaneous luminosity) for the muon of largest \pt (leading muon),
and $8\GeV$ for the subleading muon. For electrons, the corresponding trigger thresholds are $17\GeV$ and $8\GeV$.
Offline, muon candidates are reconstructed through a simultaneous fit to the hits recorded in the tracker and the muon detectors~\cite{Chatrchyan:2012xi}.
Electrons are reconstructed using both calorimeter and tracking information~\cite{CMS-PAS-EGM-10-004}.
The two leptons of largest \pt (i.e. the two leading leptons) in the event are required to be of opposite electric charge and have $\pt>20\GeV$, $|\eta|<2.4$, and invariant mass
satisfying $71< m_{\ell\ell} < 111\GeV$ to be considered \Z boson candidates.
The lepton candidates are also  required to be isolated from other energy depositions in the event.
In particular, an isolation variable is computed using the scalar sum of transverse momenta of tracks and calorimetric energy depositions
within a cone defined by $\Delta R=\sqrt{(\Delta \phi)^2 + (\Delta \eta)^2}=0.3$
around the direction of the lepton, where $\Delta\phi$ is in radians.
The contribution from pileup to this \pt sum is estimated from the distribution of the energy
per unit area in the $\eta$-$\phi$ plane in each event~\cite{Cacciari:2007fd}, and is subtracted from the calculated sum.
This corrected sum is required to be less than 15\% of the measured \pt of the lepton.
Lepton reconstruction efficiencies are determined using simulation, and corrected for differences between data and simulation using the ``tag-and-probe'' technique described in Ref.~\cite{inclusiveWZ}.

The inputs to the CMS jet clustering algorithm are the four-momentum vectors of the particles reconstructed using the particle-flow (PF) technique~\cite{CMS-PAS-PFT-09-001,CMS-PAS-PFT-10-002},
which combines information from different subdetectors.
Jets are reconstructed using the anti-\kt clustering algorithm~\cite{Cacciari:2008gp}, with a size parameter of $R=0.5$, by summing the four-momenta of individual PF particles according to the
\fastJET package of Refs.~\cite{fastjet1,fastjet2}.

The reconstructed PF candidates are calibrated to account for any nonlinear or nonuniform response of the CMS calorimetric system to neutral hadrons.
Charged hadrons and photons are sufficiently well-measured in the tracker and in the ECAL, and do not need such corrections. However,
the resulting jets require small additional energy adjustments, mostly from thresholds set on reconstructed tracks and from the clustering procedure in the PF algorithm,
but also from biases generated through inefficiencies in reconstruction. Jet-energy corrections are obtained using simulated events that are generated with \PYTHIA (version 6.4.22),
processed through a CMS detector simulation based on {\sc geant4}, and then combined with measurements of exclusive two-jet and photon~+~jet events from data~\cite{JES2011}.
By design, jet energy corrections bring reconstructed jets from detector level to particle level~\cite{Buttar:2008jx}, as opposed to the parton level.
An offset correction is also applied to account for the extra energy clustered in jets from the presence of additional proton-proton interactions (in-time or out-of-time pileup)
within the same or neighboring bunch crossings.
The overall jet-energy corrections depend on the $\eta$ and \pt values of jets, and
are applied as multiplicative factors to the four-momentum vector of each jet.
These factors range between 1.0 and 1.2, and are approximately uniform in $\eta$.
The jets accepted for analysis are required to satisfy $\pt>50\GeV$ and $|\eta|<2.5$.
In addition, all jet axes are required to be separated by $\Delta R>0.4$ from those of lepton candidates from $\Z \rightarrow \ell^+ \ell^-$ decays.
From MC studies, it is found that the selection efficiency of \Z + jets candidates is almost independent of jet multiplicity.

\section{Observable quantities}

\begin{figure}[hbt]
  \begin{center}
    \setsubfloat{0.4\textwidth}{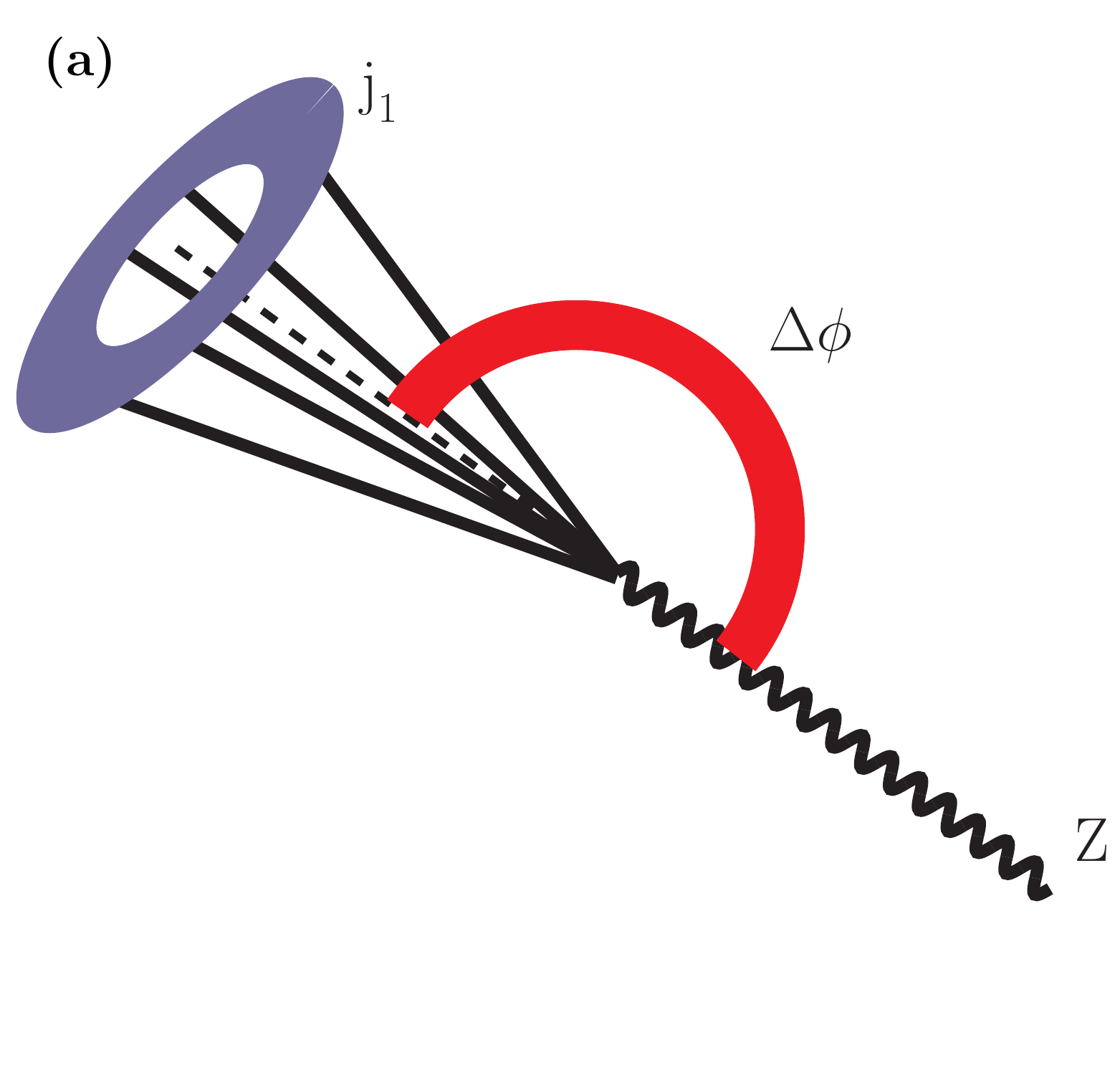}{fig:cartoon:oneJet}
    \setsubfloat{0.4\textwidth}{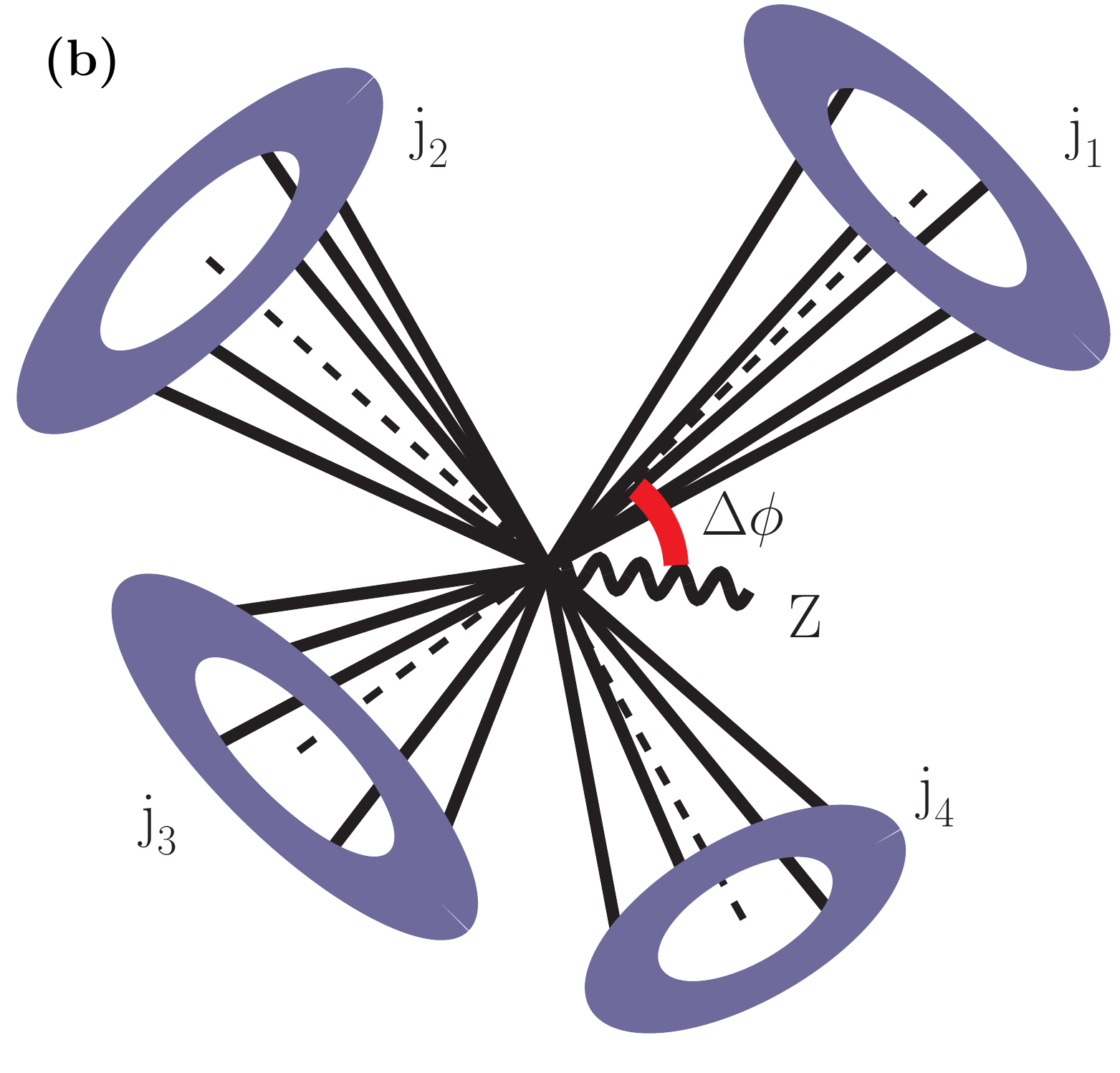}{fig:cartoon:threeJet}
    \caption{Topology of \Z + jets events:
      (a) for $\ln\tau_\mathrm{T}\rightarrow -\infty$ and $\Delta\phi(\cPZ,j_1)\rightarrow \pi$;
      (b) for $\ln\tau_\mathrm{T}\rightarrow -1$ and $\Delta\phi(\cPZ,j_1) \ll \pi$.}
    \label{fig:cartoon}
  \end{center}
\end{figure}

The observable quantities used to describe the properties of \zjets events are the differential cross sections as functions of the azimuthal angles $\Delta\phi(\cPZ,j_{i})$ between the
transverse-momentum vectors of the \Z boson and the ${i}^{\mathrm{th}}$ leading jet in the event;
the azimuthal angles among the three jets of leading \pt $\Delta\phi(j_{i},j_{k})$, with $i<k$, and $i$ and $k$ corresponding to $1, 2,$ or $3$;
and the transverse thrust $\tau_\mathrm{T}$, defined as~\cite{evshpSalam}

\begin{equation}\label{eq:thrust}
\tau_\mathrm{T}\equiv 1-\max_{\vec{n}_{\tau}}\frac{\sum_{i}\left|\vec{p}_{\mathrm{T},i}\cdot \vec{n}_{\tau} \right|}{\sum_{i}p_{\mathrm{T},i}},
\end{equation}

where the four-momenta of the \Z boson and the jets are used as inputs to calculate $\tau_\mathrm{T}$, with $\vec{p}_\mathrm{T,i}$ being the transverse-momentum vector of object ${i}$, and the sum
running over the \Z and each accepted jet in the event.
The unit vector $\vec{n}_{\tau}$ that maximizes the sum, and thereby minimizes $\tau_\mathrm{T}$, is called the thrust axis.
In the limit of the production of back-to-back \Z + 1-jet events, $\tau_\mathrm{T}$ tends to zero (Fig.~\ref{fig:cartoon}a).
With additional jet emission (i.e. the appearance of a second jet), the values of thrust increase.
Thrust is most sensitive to specifics of modeling of two-jet and three-jet topologies, while it is less sensitive to QCD modeling of larger jet multiplicities.
For clarity of presentation, we display results in terms of $\ln\tau_\mathrm{T}$ rather than $\tau_\mathrm{T}$.
The largest possible value is reached in the limit of a spherical, isotropically distributed event, where  $\ln\tau_\mathrm{T}\to \ln(1-2/\pi) \approx -1$ (Fig.~\ref{fig:cartoon}b).
In this limiting case the term $2/\pi$ originates from the uniform azimuthal distribution of the transverse momenta.

\begin{figure*}[hbt]
  \begin{center}
    \includegraphics[width=0.49\textwidth]{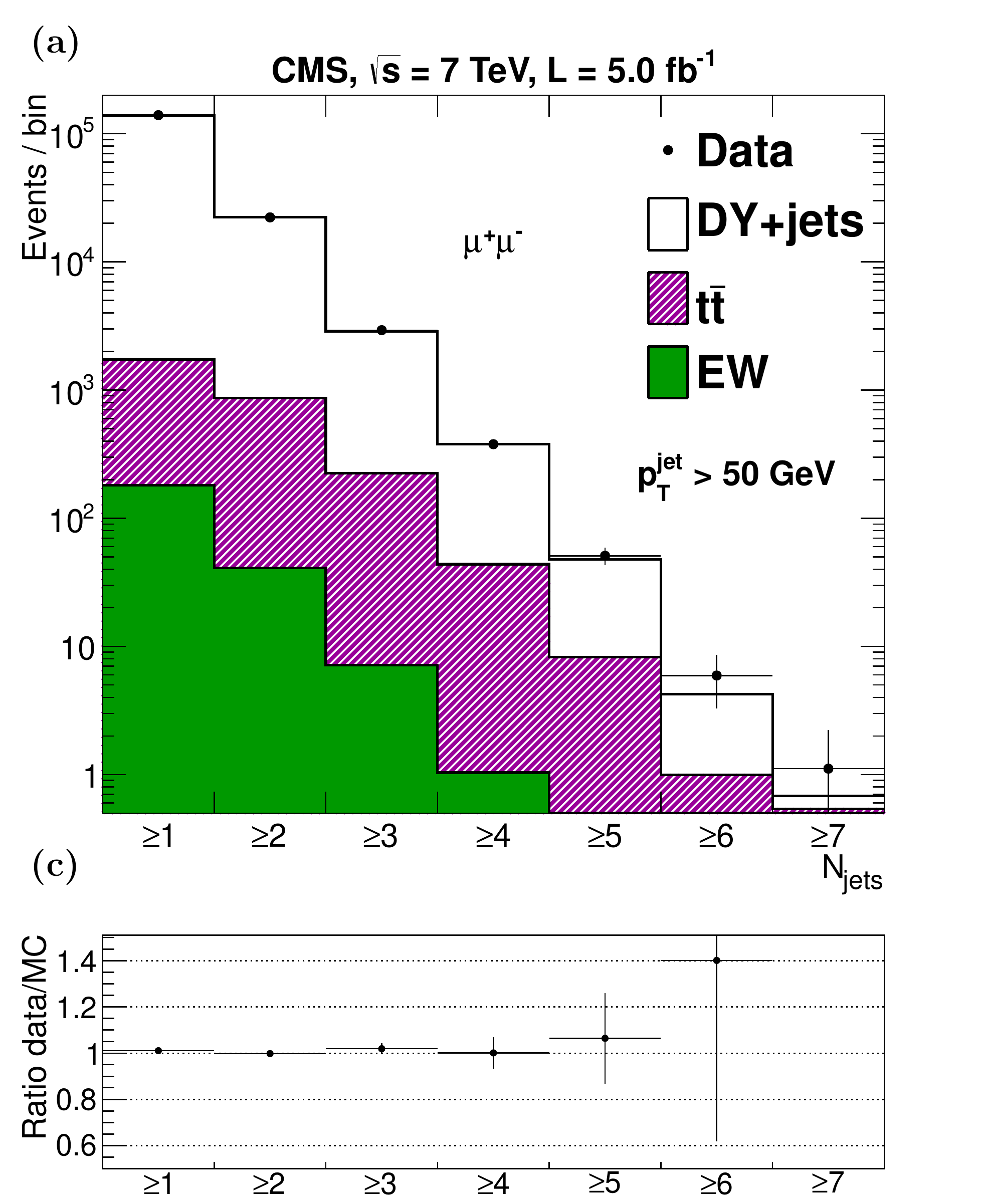}
    \includegraphics[width=0.49\textwidth]{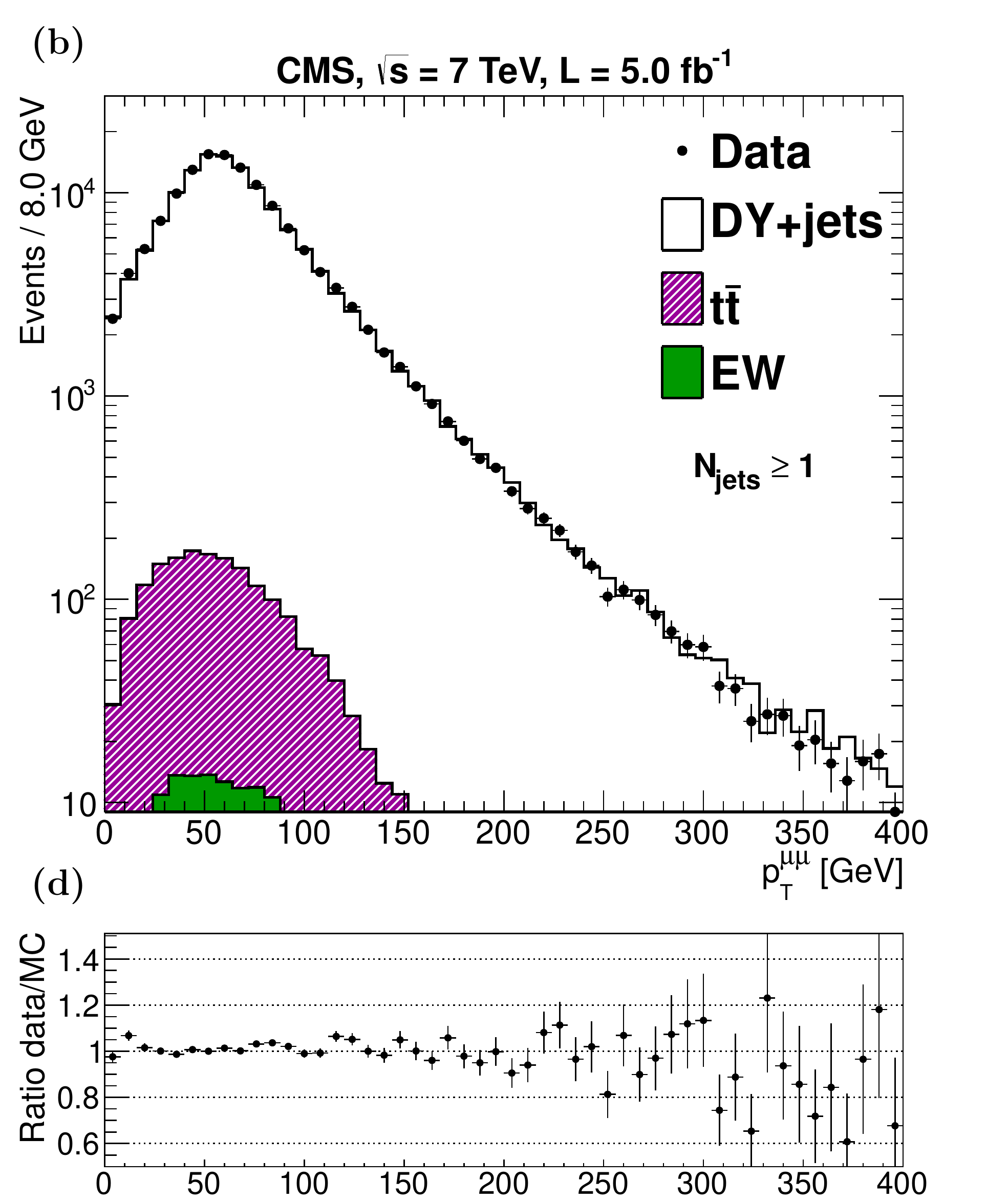}
    \caption{
Distributions for \Zmumu candidate events in data, compared with expectations from simulated signal and background contributions using \MADGRAPH simulations normalized to the
integrated luminosity of the data:
(a) as a function of associated jet multiplicity $\njets$, and (b) as a function of \pt of the dimuon pair \mbox{\ensuremath{(\pt^{\mu\mu})}\xspace} for \ensuremath{\njets\geq 1}.
The dibosons \WW, \WZ, \ZZ and \W + jets backgrounds are collectively denoted as EW in the legends.
The plots in (c) and (d) show the ratios of the data to predictions from \MC.
The error bars on the data points represent only their statistical uncertainties and do not include systematic effects.
}
    \label{fig:genkin}
  \end{center}
\end{figure*}

\begin{figure*}[hbt]
  \begin{center}
    \includegraphics[width=0.49\textwidth]{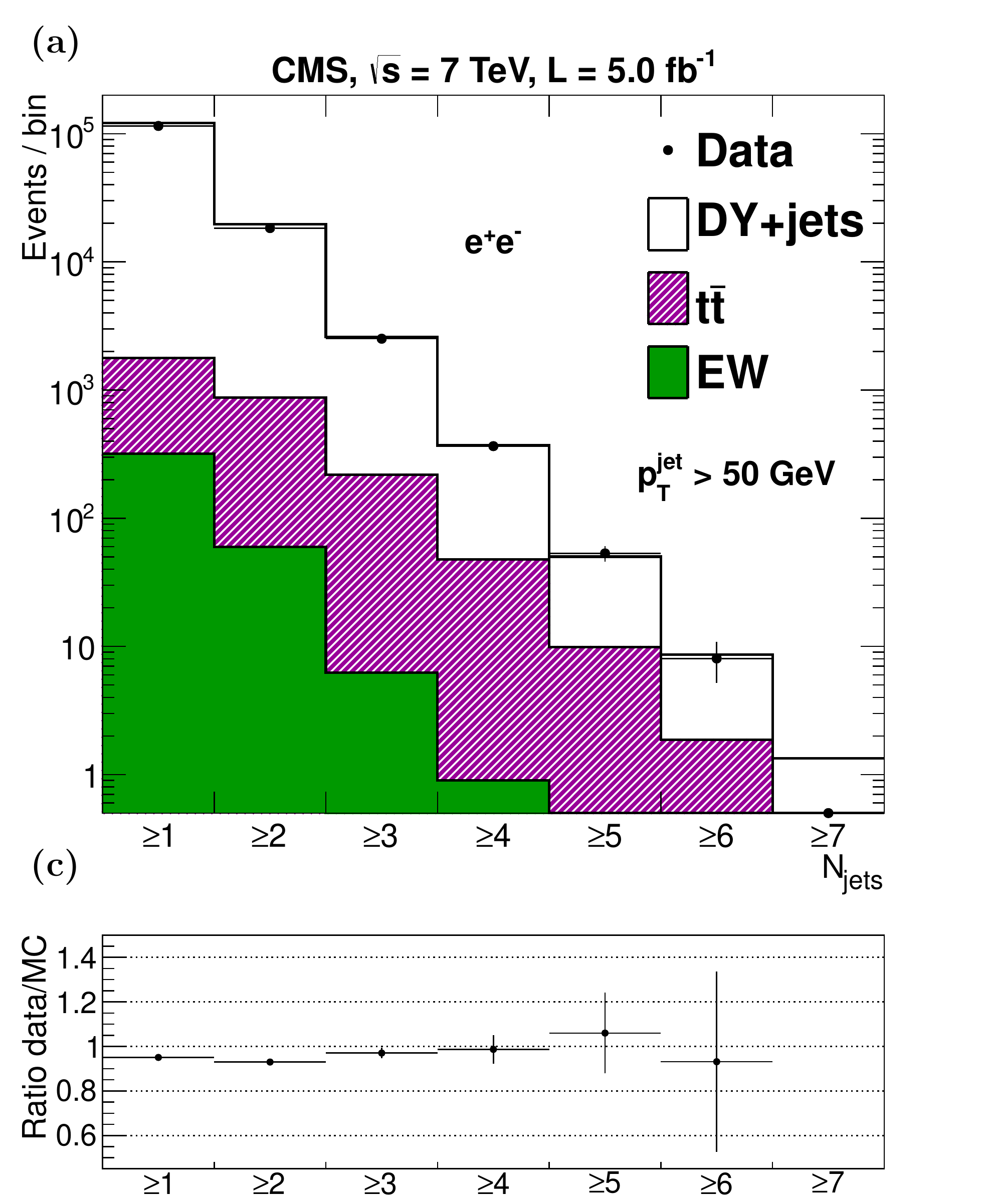}
    \includegraphics[width=0.49\textwidth]{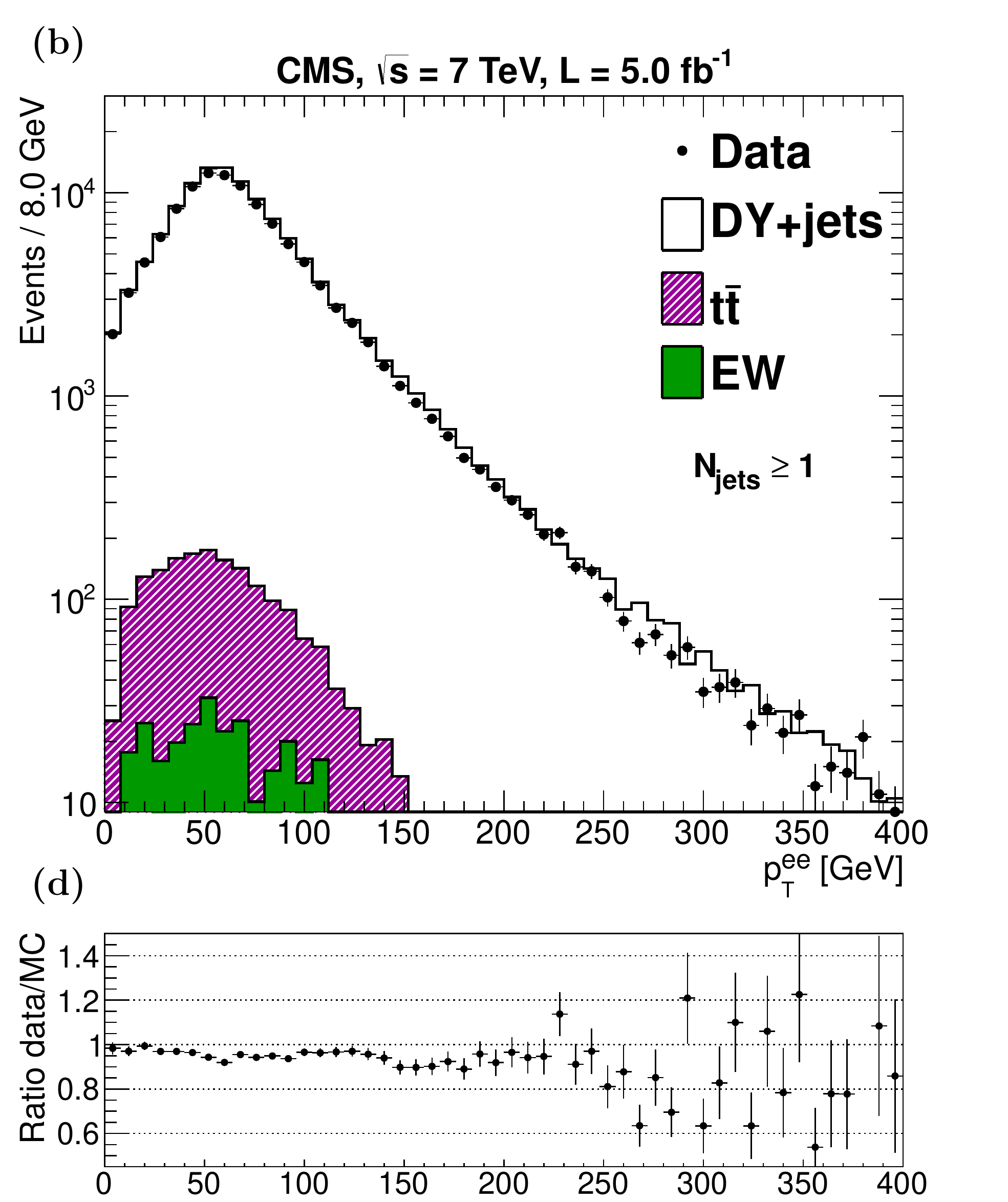}
    \caption{
Distributions for \Zee candidate events in data, compared with expectations from simulated signal and background contributions using \MADGRAPH simulations normalized to the
integrated luminosity of the data:
(a) as a function of associated jet multiplicity $\njets$, and (b) as a function of \pt of the dielectron pair \mbox{\ensuremath{(\pt^{ee})}\xspace} for \ensuremath{\njets\geq 1}.
The dibosons \WW, \WZ, \ZZ and \W + jets backgrounds are collectively denoted as EW in the legends.
The plots in (c) and (d) show the ratios of the data to predictions from \MC.
The error bars on the data points represent only their statistical uncertainties and do not include systematic effects.
}
    \label{fig:genkin_e}
  \end{center}
\end{figure*}

To investigate the dependence of the topological properties on the complexity of the final state, the events are categorized as a function of jet multiplicity.
In particular, the azimuthal distributions are reported in inclusive bins of one, two, or three jets.
Furthermore, the phase space is characterized according to the \pt of the \Z boson, and measurements performed either for all $\ptZ$ or in the region of $\ptZ>150\GeV$.
Figure~\ref{fig:genkin} and ~\ref{fig:genkin_e} show, respectively, the distributions in the associated jet multiplicity and the \pt
of $\Pgmp\Pgmm$ and $\Pep\Pem$ systems for $\geq$1 jet events,
prior to background subtraction.
Both sets of distributions are presented at the detector level, and are within statistical uncertainty of the \MC predictions for DY+jets, \ttbar, and other electroweak (EW) background sources.
It should be noted that $\ptZ$ refers to the transverse momentum of the \Z boson, following background subtraction and unfolding of detector effects.

\section{Analysis procedure and systematic uncertainties}

The analysis procedure consists of the following steps:  \Zee and \Zmumu candidates are selected as described in Section~\ref{eventreco}; the background is then subtracted and
the resulting distributions are unfolded to the particle level; finally, the two channels are combined.
The dominant sources of systematic uncertainty arise from uncertainties in jet-energy scale, resolution of jet \pt, background subtraction, and the unfolding procedure.
Individual steps of the analysis procedure are detailed below.

The \zjets candidates include several sources of SM~background (Fig.~\ref{fig:genkin}), which are subtracted using predictions of the \MADGRAPH \MC event generator.
The dominant background is from \ttbar production, and is about 1.1\%, 4\%, and 8\%, for the $\njets \geq 1$, $\geq 2$, and  $\geq 3$ inclusive bins of jet multiplicity.
An independent evaluation of the background from \ttbar events is also obtained from an $\Pe\mu$ control sample in data, which is selected by requiring the presence of an electron and a muon of
opposite charge, but otherwise using the same criteria as used for selecting the \zjets events.
For each jet-multiplicity bin, the estimates obtained from data and \MC simulation agree within 6\%.
The two estimates are consistent, given the uncertainties on the integrated luminosity (2.2\%)~\cite{CMS-PAS-SMP-12-008} and on the \ttbar cross section (6\%)~\cite{ttbarxsec}.
Background originating from dibosons is $0.06\%$, $0.06\%$ and $0.01\%$ for the WW, WZ and ZZ channels, respectively.
The contribution from \Ztautau is ${<}0.08\%$, largely because of the requirement on the Z mass of $71{<}m_{\ell\ell}{<}111\GeV$.
Contributions from W+jets to the two-electron final state are estimated to be $0.1\%$, and ${<}0.1\%$ to the two-muon channel.
The contribution from single top-quark production is estimated as ${<}0.1\%$.
Background from multijet production, estimated using dilepton pairs of same electric charge, is found to be negligible.
A total uncertainty of 10\% is assigned to the expectation from all background sources.
The limited contribution of backgrounds to the \Z boson candidate sample is reflected in a ${<}$1\% uncertainty
on the final measurement.

The particle-level four-momentum vector of a lepton is computed in the \MC simulation by adding the four-momentum vectors of any photons found within a radius of $\Delta R=0.1$ of each lepton axis
to the four-momentum vector of the lepton.
For the observables of interest in this analysis, the use of this cone size makes the electron and muon channels essentially the same at the particle level.
In this way, the difference in final-state radiation in the \Zmumu and \Zee channels is accounted for and the two channels can then be directly combined.
The particle-level jets in MC events are reconstructed by clustering the generated stable particles (after hadronization) using the same anti-\kt algorithm, with a parameter $R=0.5$,
as done in data. The selection criteria used in data are also applied to particle-level leptons and jets: the two leading leptons are required to have $\pt>20\GeV$ and $|\eta|<2.4$,
while the jets must have $\pt>50\GeV$ and $|\eta|<2.5$. An angular separation of $\Delta R(\ell,j)>0.4$ is also required between the two leading leptons and any accepted jet.
Finally, the unfolded distributions from the \Zee and \Zmumu channels are combined at the level of covariance matrices using the best linear unbiased estimator~\cite{BLUE}.

The background-subtracted, detector-level distributions are mapped to the particle level by correcting for effects of detector resolution and efficiency.
Migration of events among bins of inclusive jet multiplicity can be caused by detector resolution, especially from the mismeasurement of jet \pt.
For example, an event containing a \Z boson produced in association with $N$ jets at the particle level, can migrate to the $N+1$ jets final state as a result of detector resolution.
The opposite effect can also occur leading to loss of events that migrate out of the geometric and kinematic acceptance.
Such migrations correspond to as much as $30\%$ and are treated in the detector unfolding procedure summarized below.
Detector effects are expressed through a response matrix, which is determined from \MC simulation, separately for each lepton flavor and each observable,
by associating the particle-level values to their reconstructed quantities.
Two alternative representations of the response matrix, one generated from \MADGRAPH (baseline) and the other from \SHERPA, are used in this procedure,
and half of the difference of the propagated results is used to define their systematic uncertainty.
The unfolding of data to the particle level is performed using the Singular Value Decomposition method~\cite{SVD}, implemented in the \textsc{RooUnfold} package~\cite{RooUnfold}.
The total systematic uncertainty due to the unfolding procedure is ${<}5\%$ for azimuthal correlations and ${<}2\%$ for the thrust analysis.

Among the azimuthal observables, the distribution of $\Delta\phi(\cPZ,j_1)$ has the largest systematic uncertainty.
This variable is particularly sensitive to the jet-energy scale, which can affect jet multiplicity, and thus the acceptance of events that enter in the $\Delta\phi(\cPZ,j_1)$ distribution.
The uncertainty on jet-energy scale varies between 1 and 3\%~\cite{JES2011}, and results in an uncertainty of 2--4\% on the
distribution in $\Delta \phi(\cPZ,j_1)$, increasing monotonically for decreasing  angles.
The impact from the resolution on \pt is estimated by changing jet resolutions by $\pm 10 \%$ (corresponding to about one standard deviation)~\cite{JES2011}, and comparing the
unfolding  correction before and after these changes.
This yields a dependence of $\approx$1\% in the normalized azimuthal distributions.
The uncertainty from pileup is estimated by changing in simulation the number of generated minimum-bias interactions by $\pm 5\%$.
The resulting uncertainty is 4\% for $\Delta \phi (Z,j_1)\approx 0$, which decreases to a negligible uncertainty for $\Delta \phi (Z,j_1)\approx \pi$.
The overall systematic uncertainty on $\Delta\phi(\cPZ,j_1)$ is about $5$--$6\%$ at values of $\Delta\phi(\cPZ,j_1) \approx 0$ and about 2\% at values close to $\pi$.

The dominant systematic uncertainty on the thrust distribution,
which corresponds to about 2\%,
is from the uncertainty in jet-energy scale, and can be understood as follows.
When the energy scale is increased, more jets enter the two sums in Eq.~\eqref{eq:thrust}, and both sums tend to shift to larger values.
Conversely, when the jet-energy scale is decreased, their values decrease.
The contribution from uncertainty in jet-energy resolution is found to affect the transverse thrust by 1\%, and
the uncertainties from selection efficiencies are ${<}2\%$ for the entire range of $\ln\tau_\mathrm{T}$, while
the uncertainties from pileup and background subtraction have negligible impact.
The first conclusion is implied in Eq.~\eqref{eq:thrust}, as soft additional pileup energy added to the hard jets
contributes simultaneously to both the numerator and denominator and, to first order, cancels in the ratio.
The second conclusion follows from the fact that the transverse thrust is measured in the inclusive \Z + ${\geq}1$-jet sample, where the signal purity is almost 99\%,
and background subtraction has therefore only a minimal impact on the measurement of $\ln\tau_\mathrm{T}$.

For $\ptZ>150\GeV$ the uncertainties on the azimuthal variables and on $\ln\tau_\mathrm{T}$ are evaluated following the same procedure as described above.
However, in addition to the uncertainties originating from previously discussed effects, the statistical limitations of the \MC samples become important
and the systematic uncertainty on the result therefore increases.
The impact of the uncertainties on the electron and muon selection efficiencies due to energy scale, trigger and resolution has been assessed and found to be negligible.

\section{Results}

\begin{figure*}[hbtp]
  \begin{center}
    \setsubfloat{0.49\textwidth}{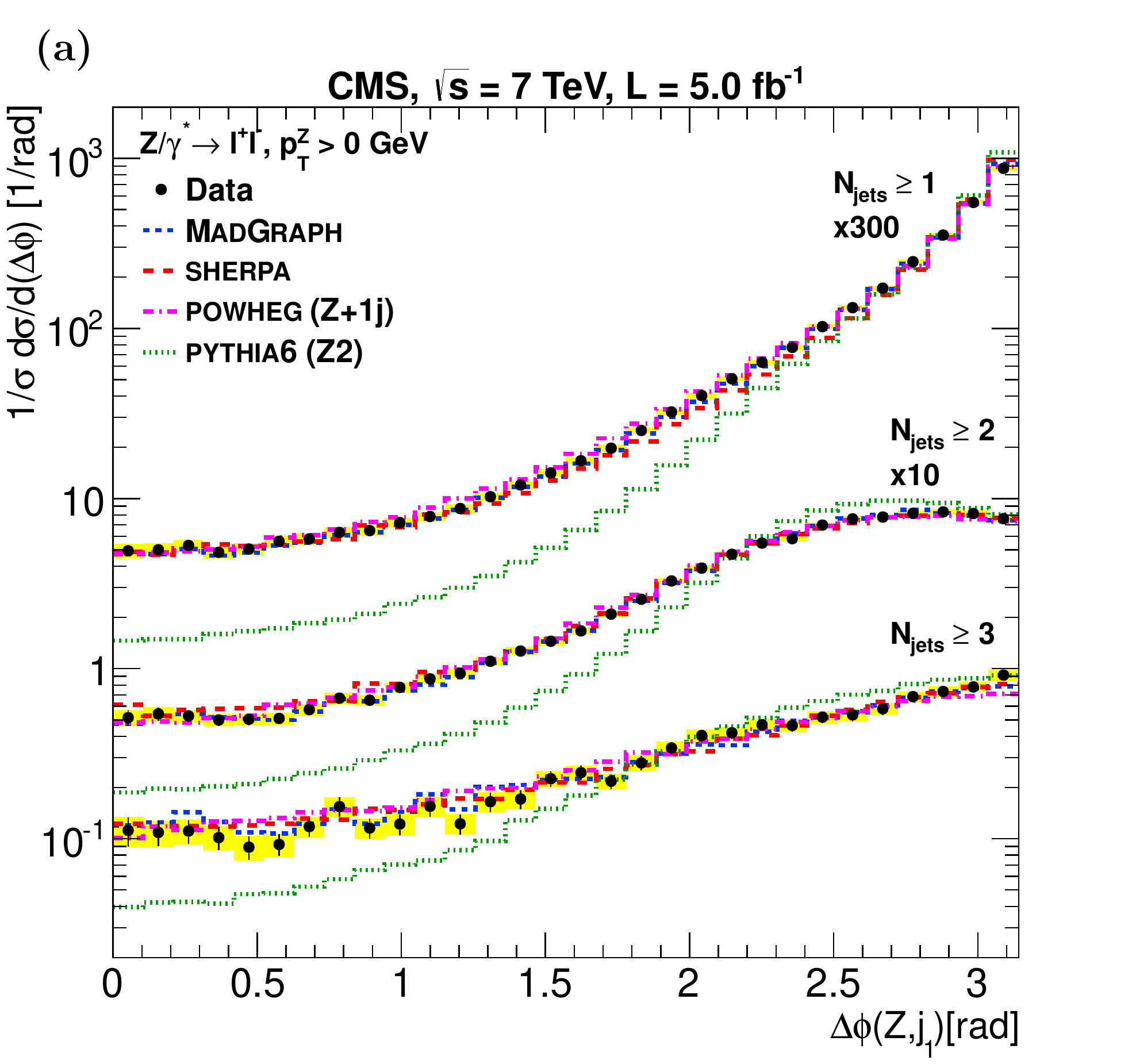}{fig:dphiZJ1jetmultbinned:Plot0}
    \setsubfloat{0.49\textwidth}{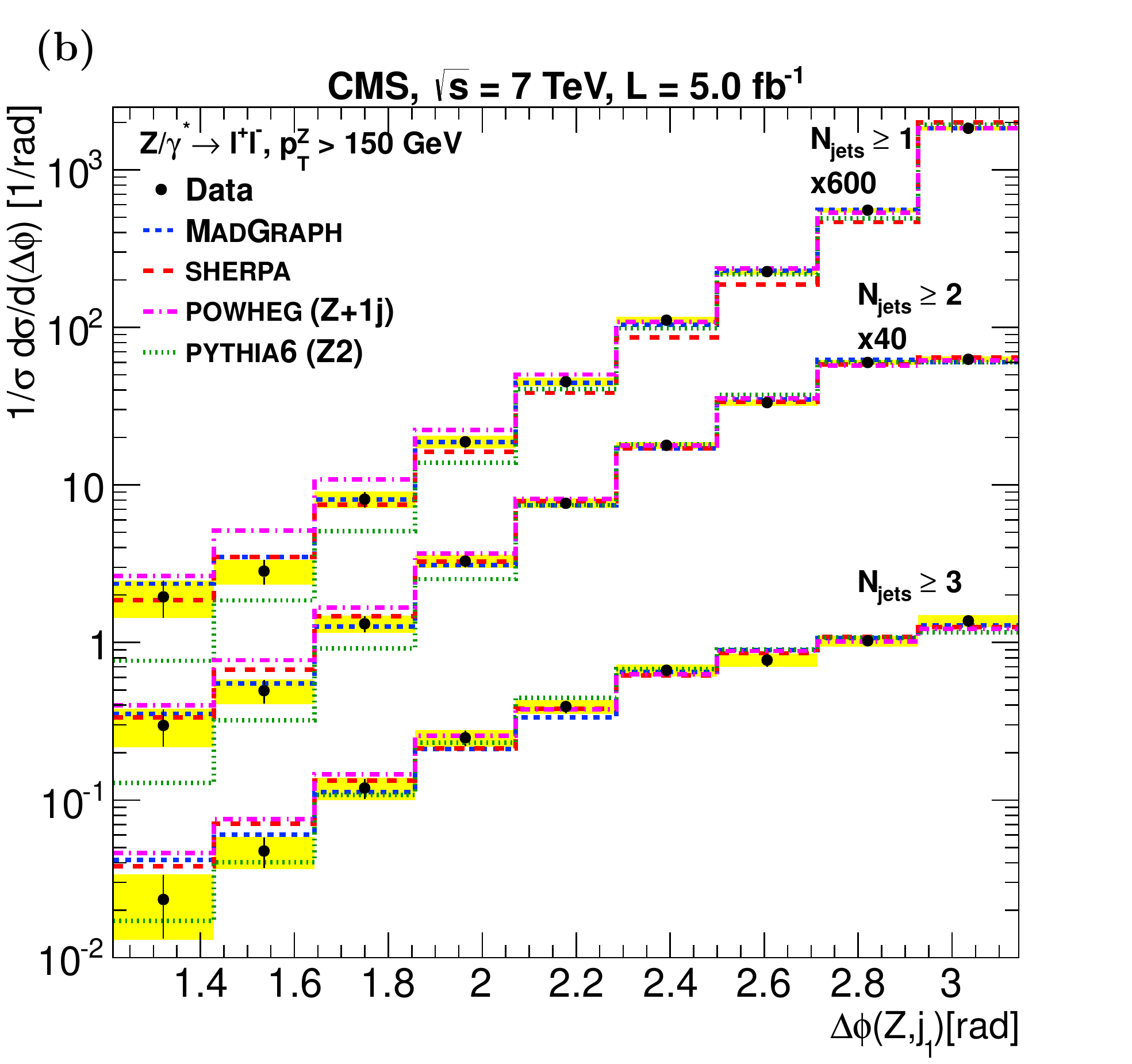}{fig:dphiZJ1jetmultbinned:Plot0Boost}
    \setsubfloatR{0.49\textwidth}{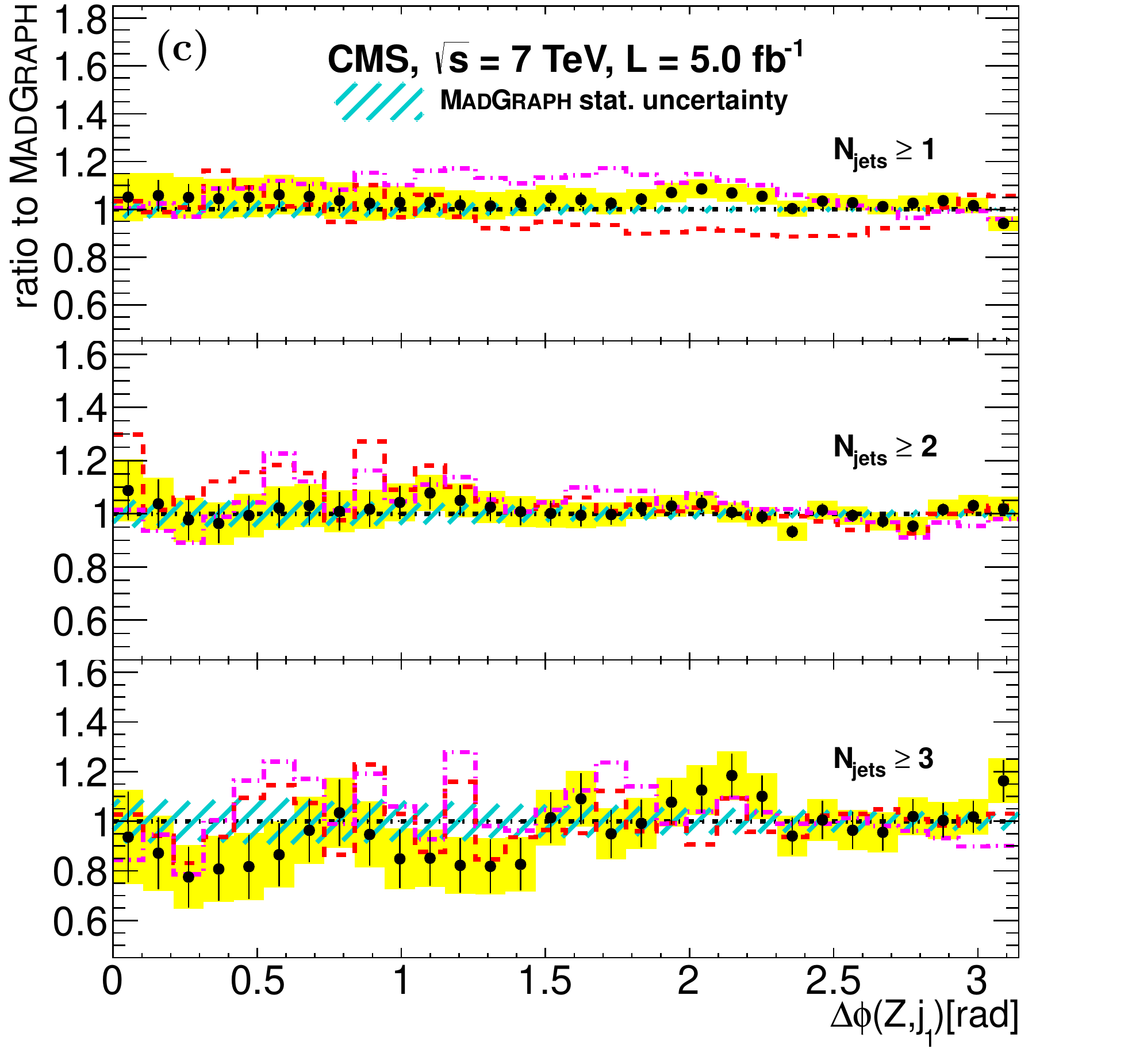}{fig:dphiZJ1jetmultbinned:Ratio0}
    \setsubfloatR{0.49\textwidth}{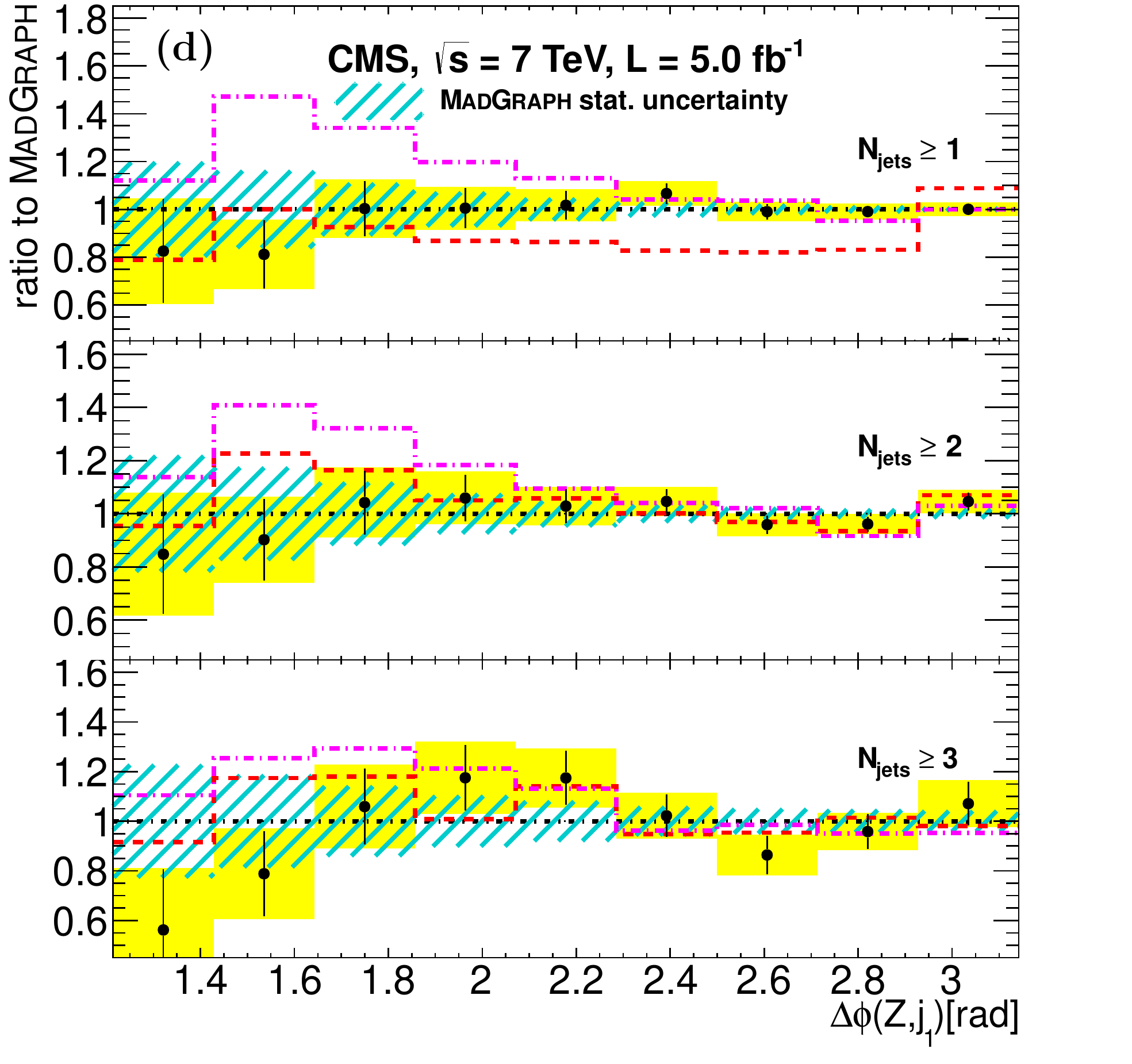}{fig:dphiZJ1jetmultbinned:Ratio0Boost}
    \caption{
Normalized $\Delta\phi(\cPZ,j_1)$ distributions for the leading jet in the inclusive jet-multiplicity bins $\njets\geq 1, \geq2,$ and $\geq3$:
(a) all \ptZ
and (b)  $\ptZ > 150\GeV$.
Plots in (c) and (d) show the corresponding ratios of the data (solid points), and of other \MC predictions,
relative to \MADGRAPH.
The ratio for \PYTHIA \MC is not included in these plots.
The error bars on the data points represent their statistical uncertainties,
the solid yellow shaded band around the points represents the sum of statistical and systematic uncertainties
taken in quadrature,
while the cross-hatched (cyan) bands reflect the statistical uncertainties on the \MADGRAPH calculations.
}
    \label{fig:dphiZJ1jetmultbinned}
  \end{center}
\end{figure*}

\begin{figure*}[hbtp]
  \begin{center}
    \setsubfloat{0.49\textwidth}{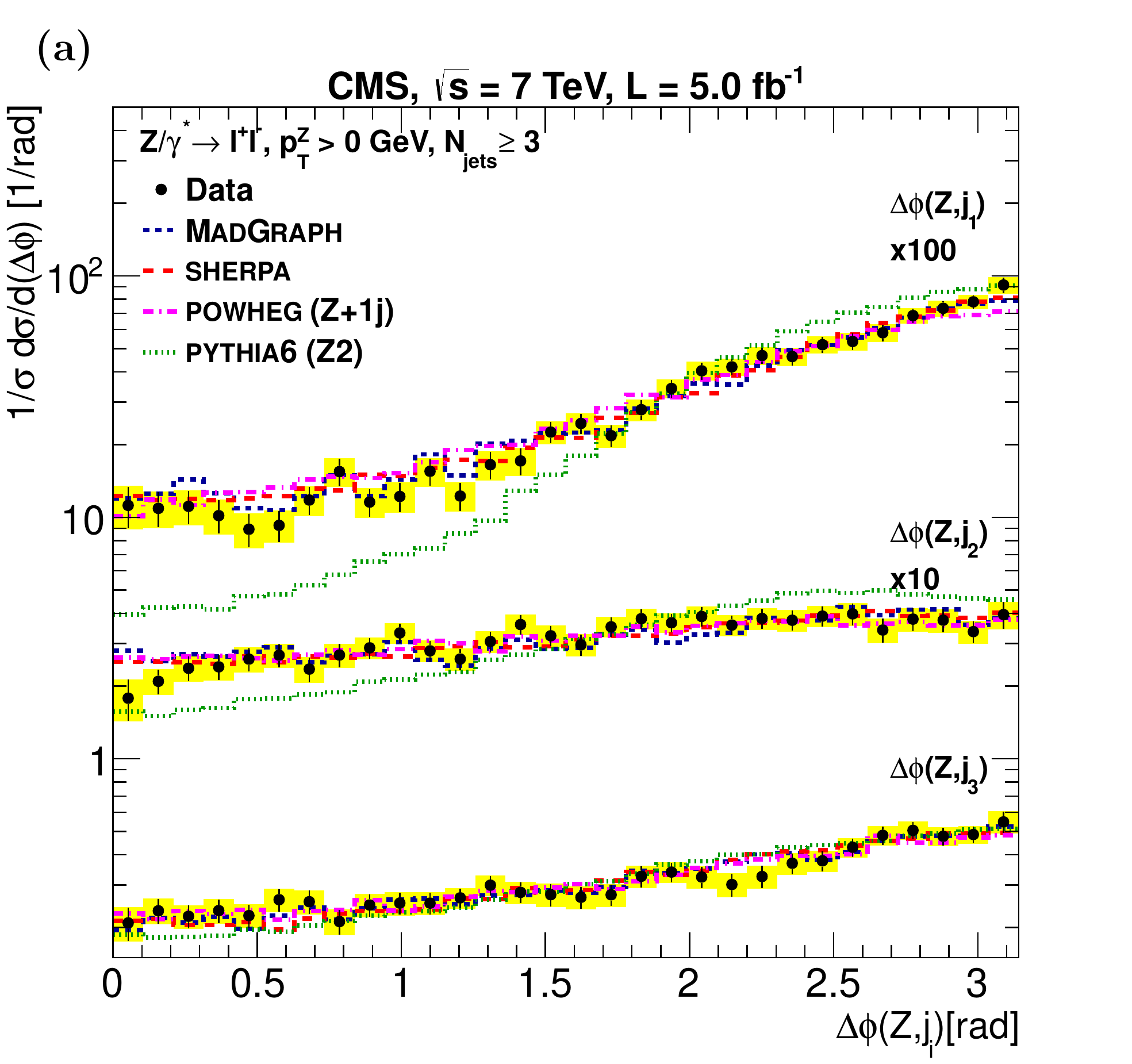}{fig:azimuth3jetspectra1:Plot1}
    \setsubfloat{0.49\textwidth}{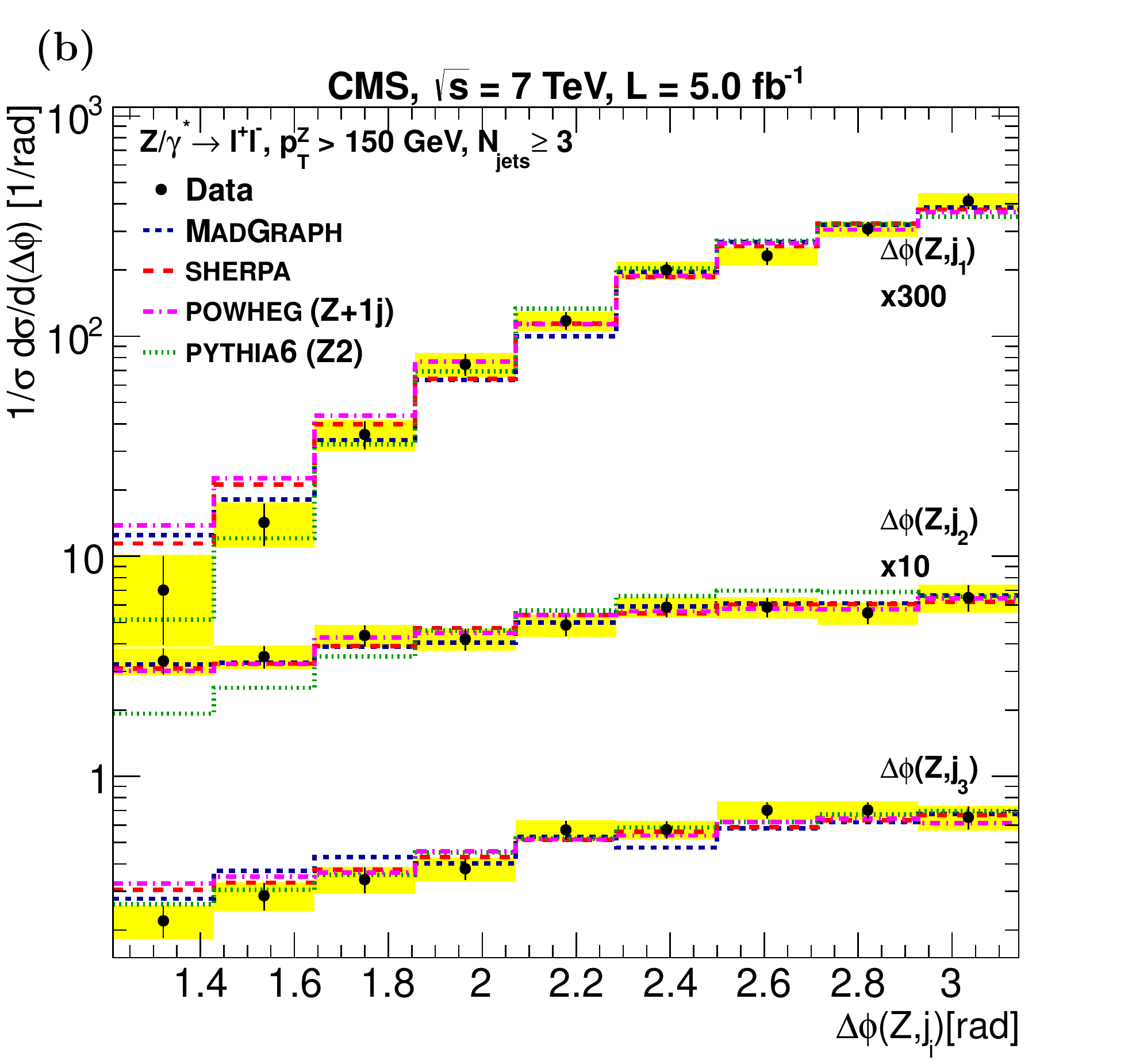}{fig:azimuth3jetspectra1:Plot1Boost}
    \setsubfloatR{0.49\textwidth}{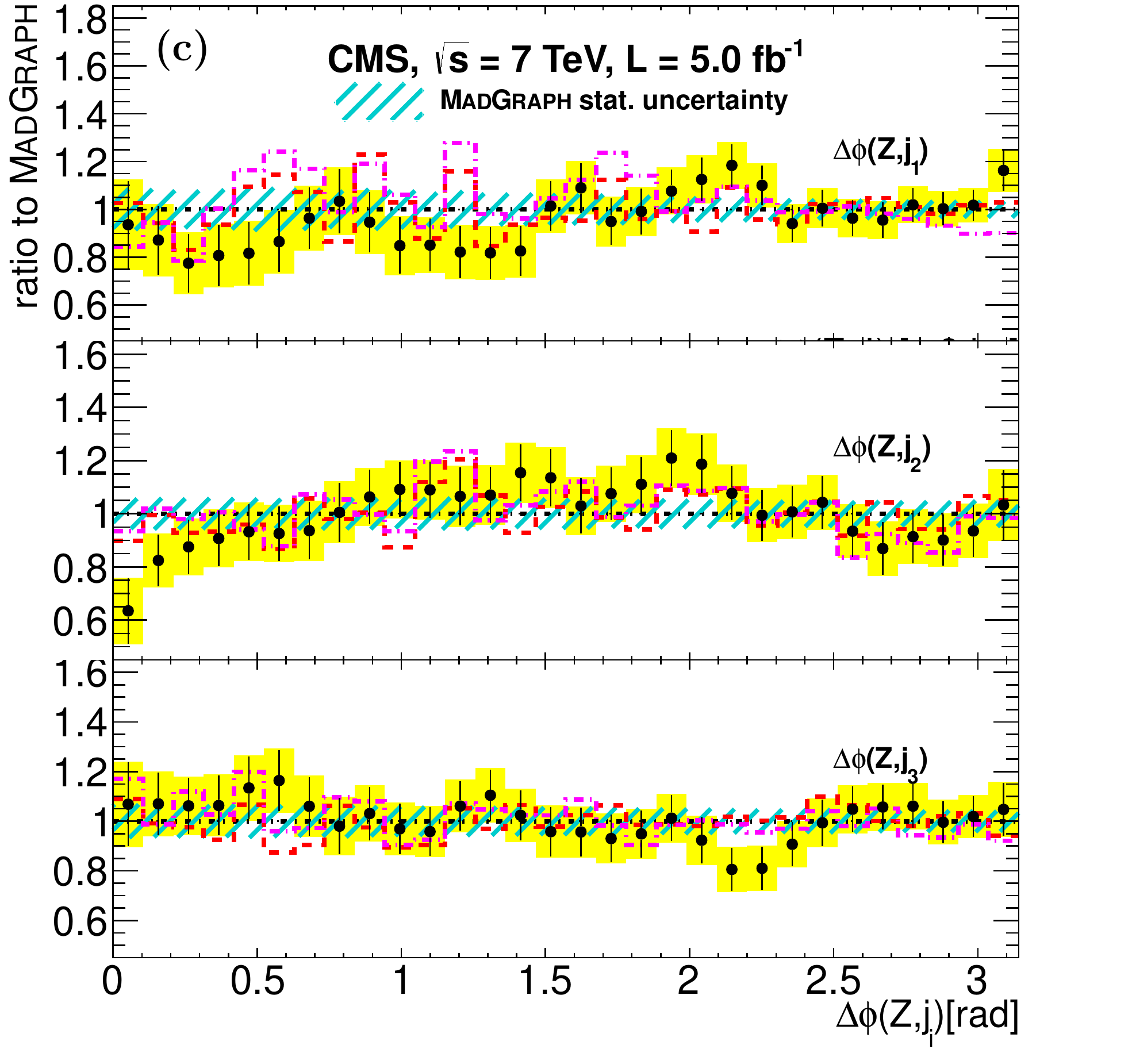}{fig:azimuth3jetspectra1:Ratio1}
    \setsubfloatR{0.49\textwidth}{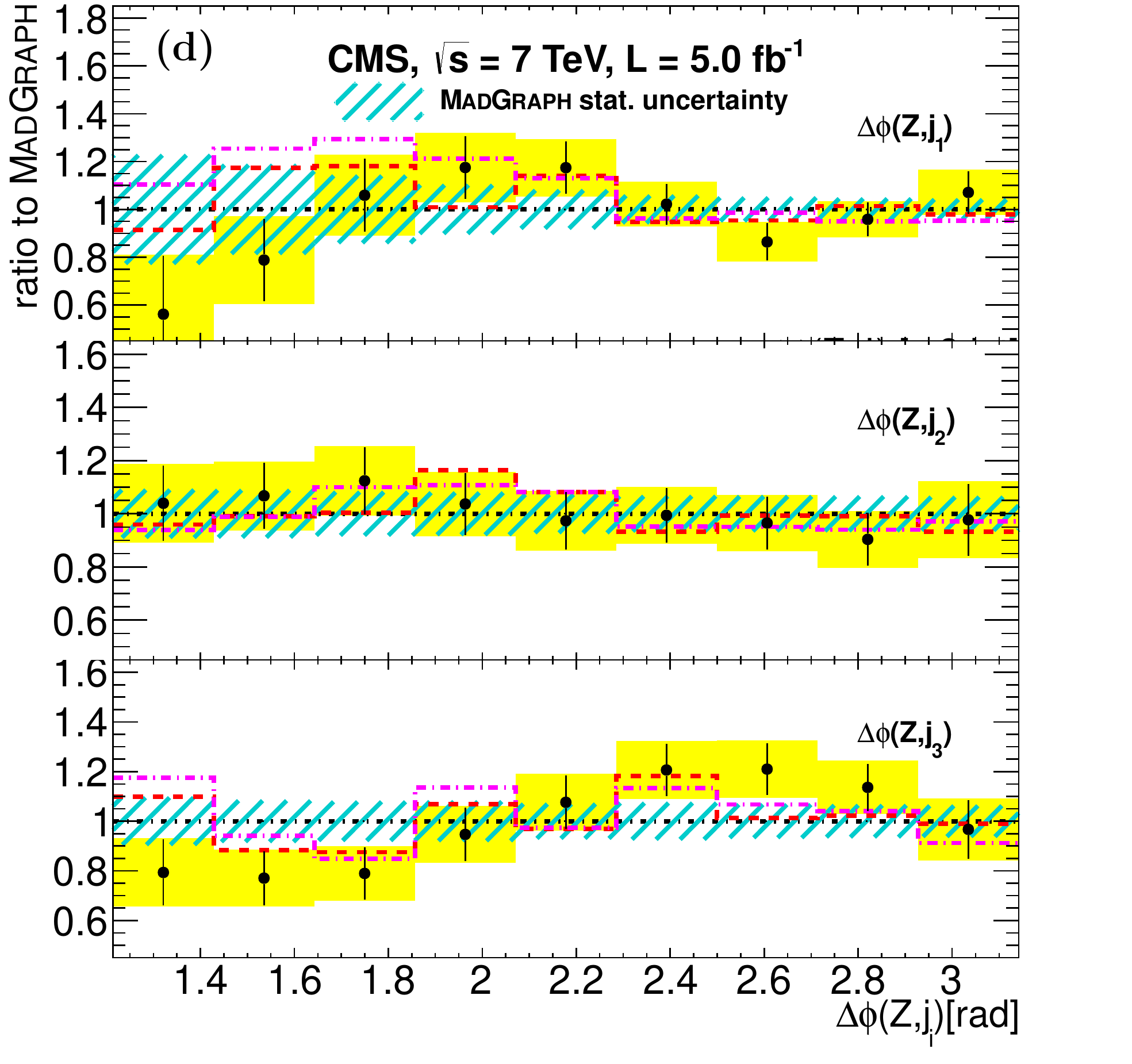}{fig:azimuth3jetspectra1:Ratio1Boost}
    \caption{
Normalized $\Delta\phi(\cPZ,j_{i})$ distributions for the inclusive $\njets\geq 3$ jet-multiplicity bin: (a)~all \ptZ
and (b)  $\ptZ > 150\GeV$.
Plots in (c) and (d) show
the ratios of the data and other \MC predictions, relative to \MADGRAPH, as described in Fig.~\ref{fig:dphiZJ1jetmultbinned}.
}
    \label{fig:azimuth3jetspectra1}
  \end{center}
\end{figure*}

\begin{figure*}[hbtp]
  \begin{center}
    \setsubfloat{0.49\textwidth}{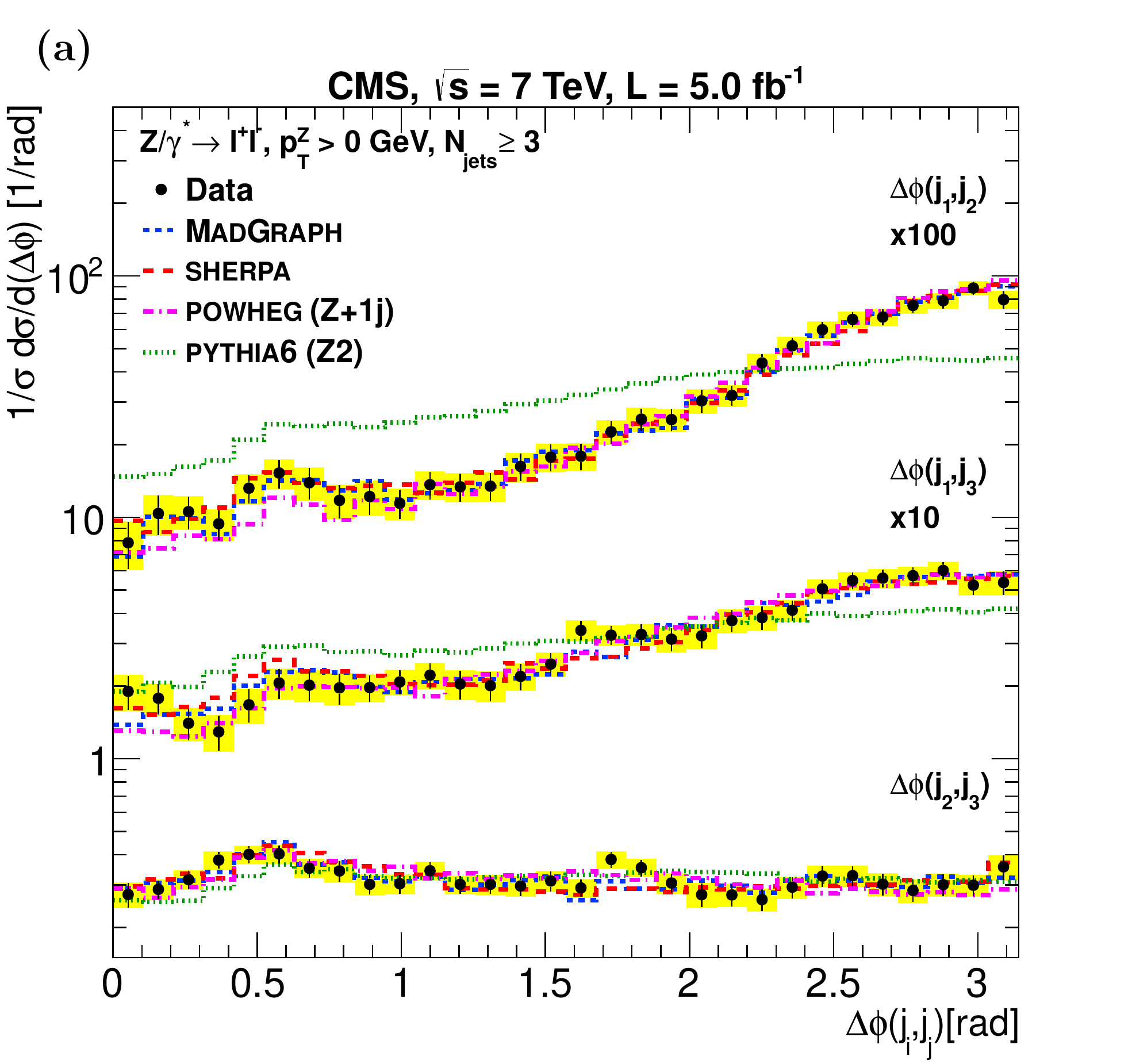}   {fig:azimuth3jetspectra2:Plot2}
    \setsubfloat{0.49\textwidth}{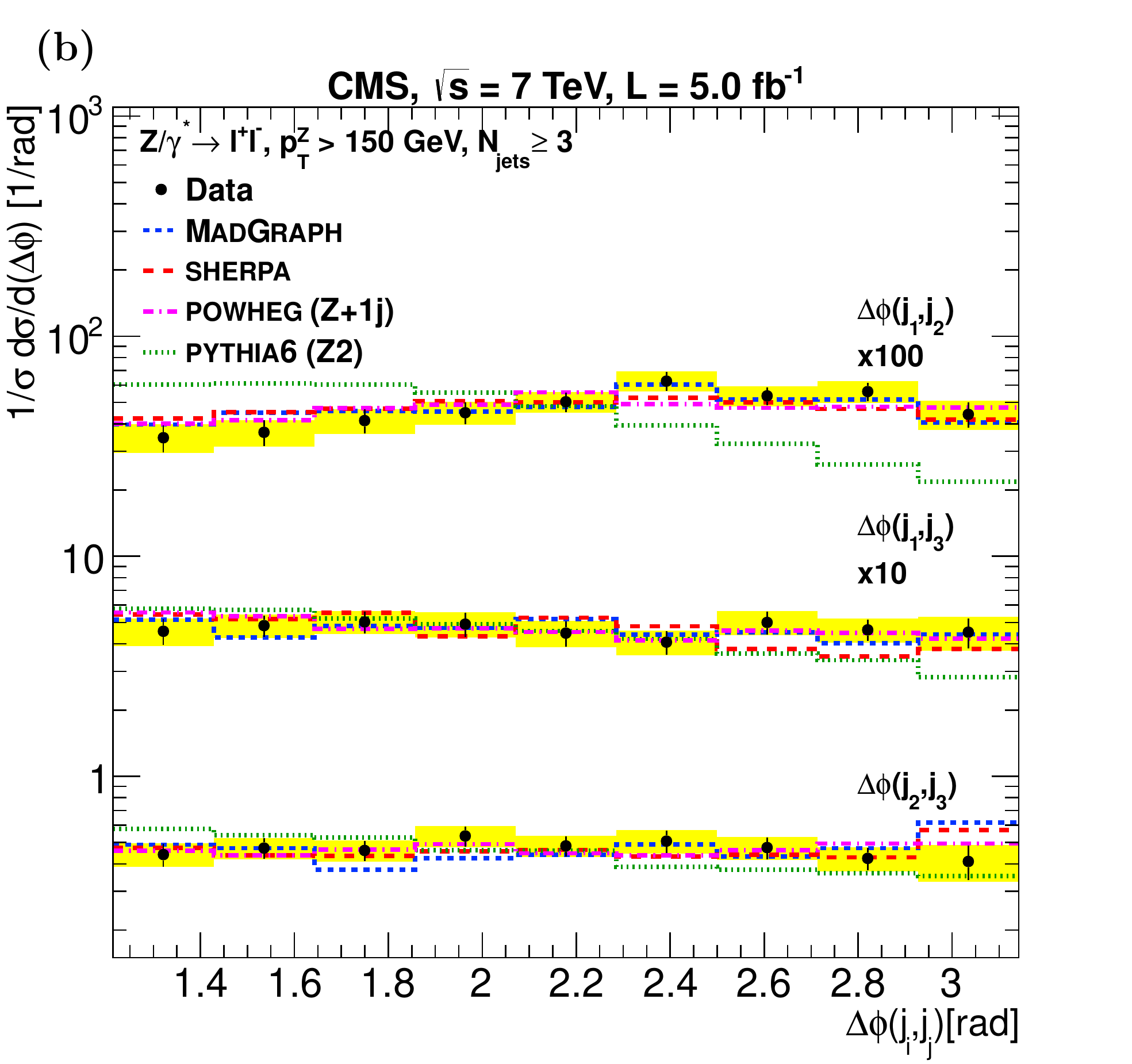} {fig:azimuth3jetspectra2:Plot2Boost}
    \setsubfloatR{0.49\textwidth}{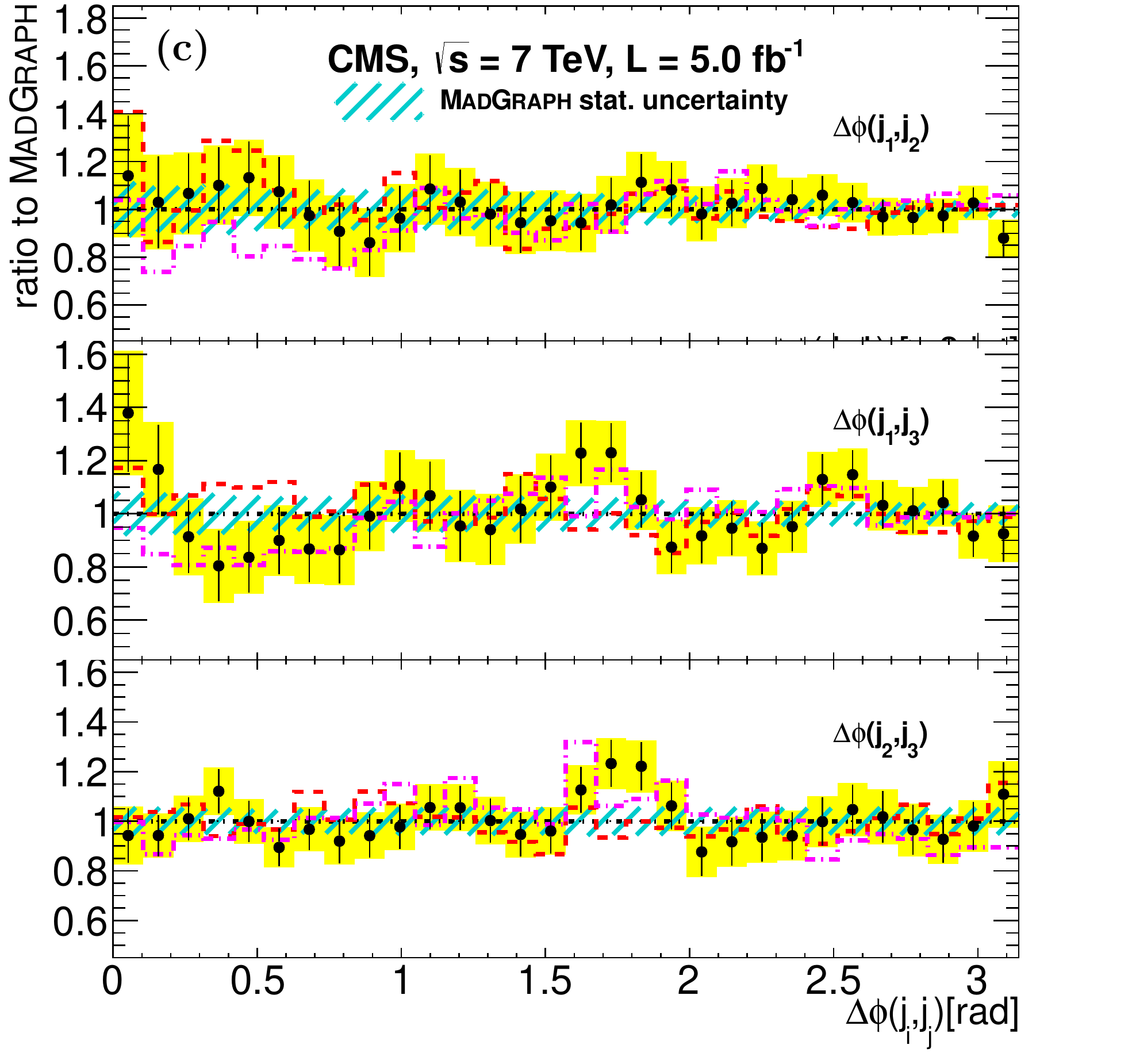}{fig:azimuth3jetspectra2:Ratio2}
    \setsubfloatR{0.49\textwidth}{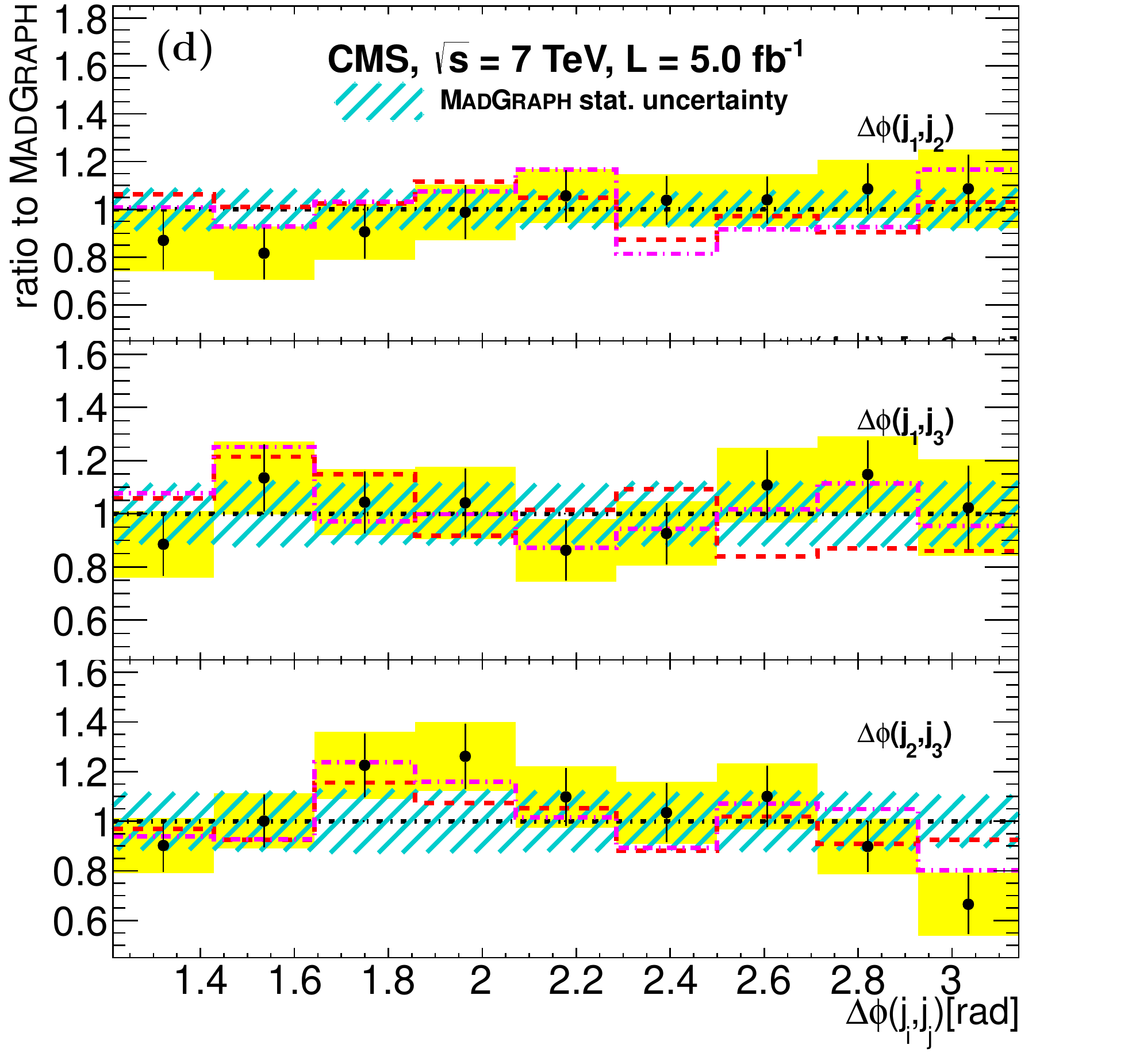}{fig:azimuth3jetspectra2:Ratio2Boost}
    \caption{
Normalized $\Delta\phi(j_{i},j_{j})$ distributions for the inclusive $\njets\geq 3$ jet-multiplicity bin: (a)~all \ptZ
and (b)  $\ptZ > 150\GeV$.
Plots in (c) and (d) show the ratios of the data and other \MC predictions, relative to \MADGRAPH, as described
in Fig.~\ref{fig:dphiZJ1jetmultbinned}.
}
     \label{fig:azimuth3jetspectra2}
  \end{center}
\end{figure*}

\begin{figure*}[hbt]
  \begin{center}
    \setsubfloatT{0.49\textwidth}{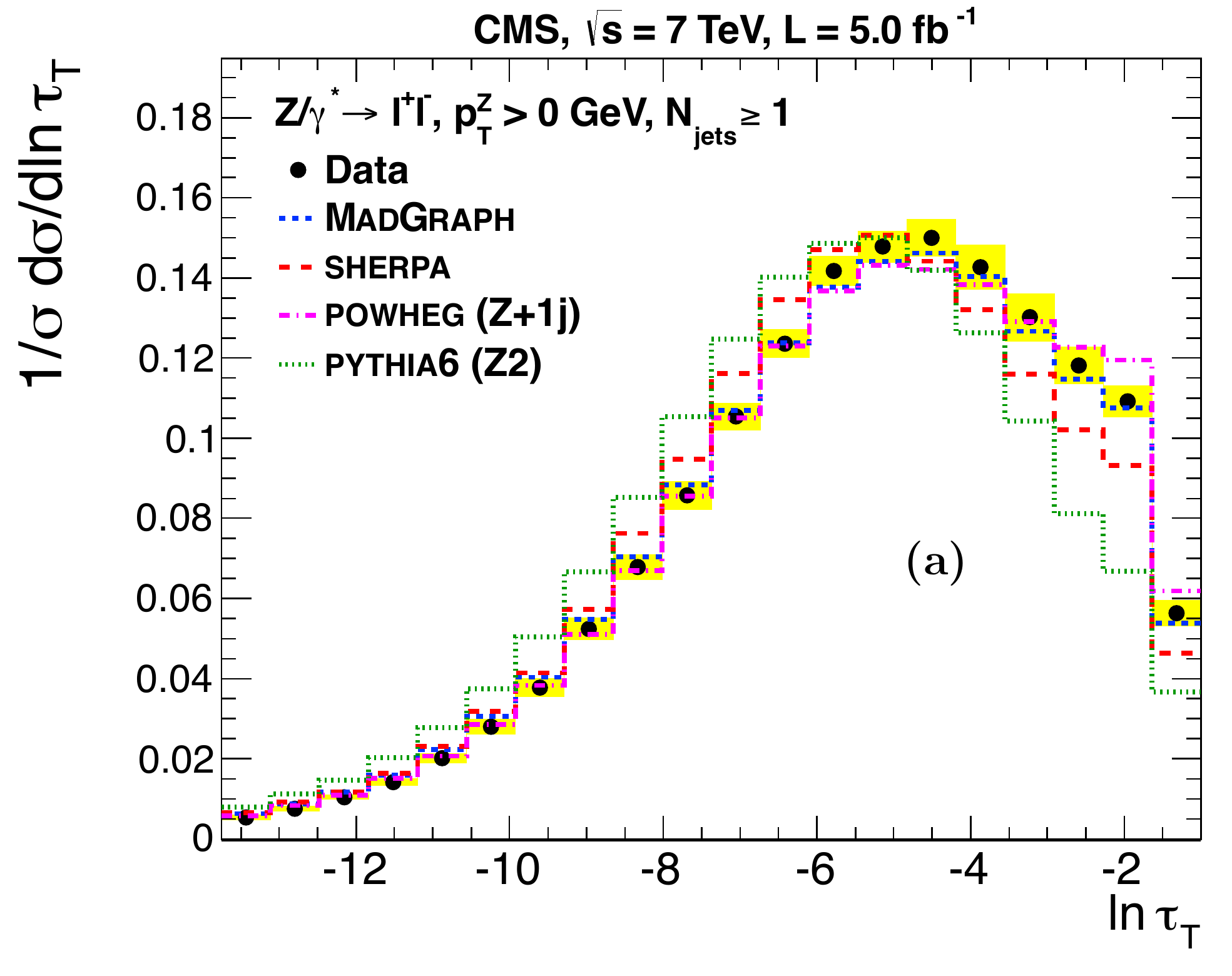} {fig:thrust_results:Plot}
    \setsubfloatT{0.49\textwidth}{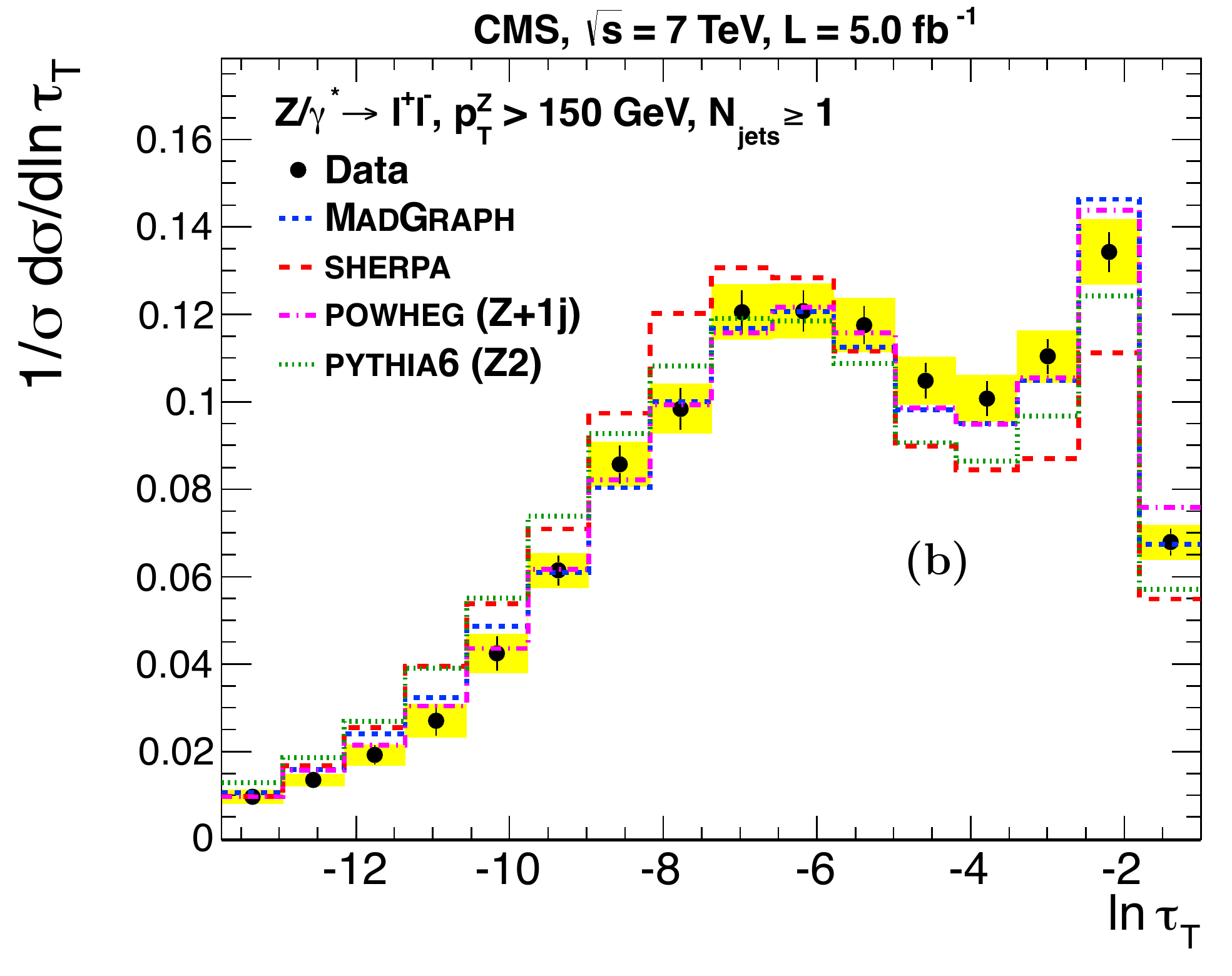}{fig:thrust_results:PlotBoost}
    \setsubfloatTR{0.49\textwidth}{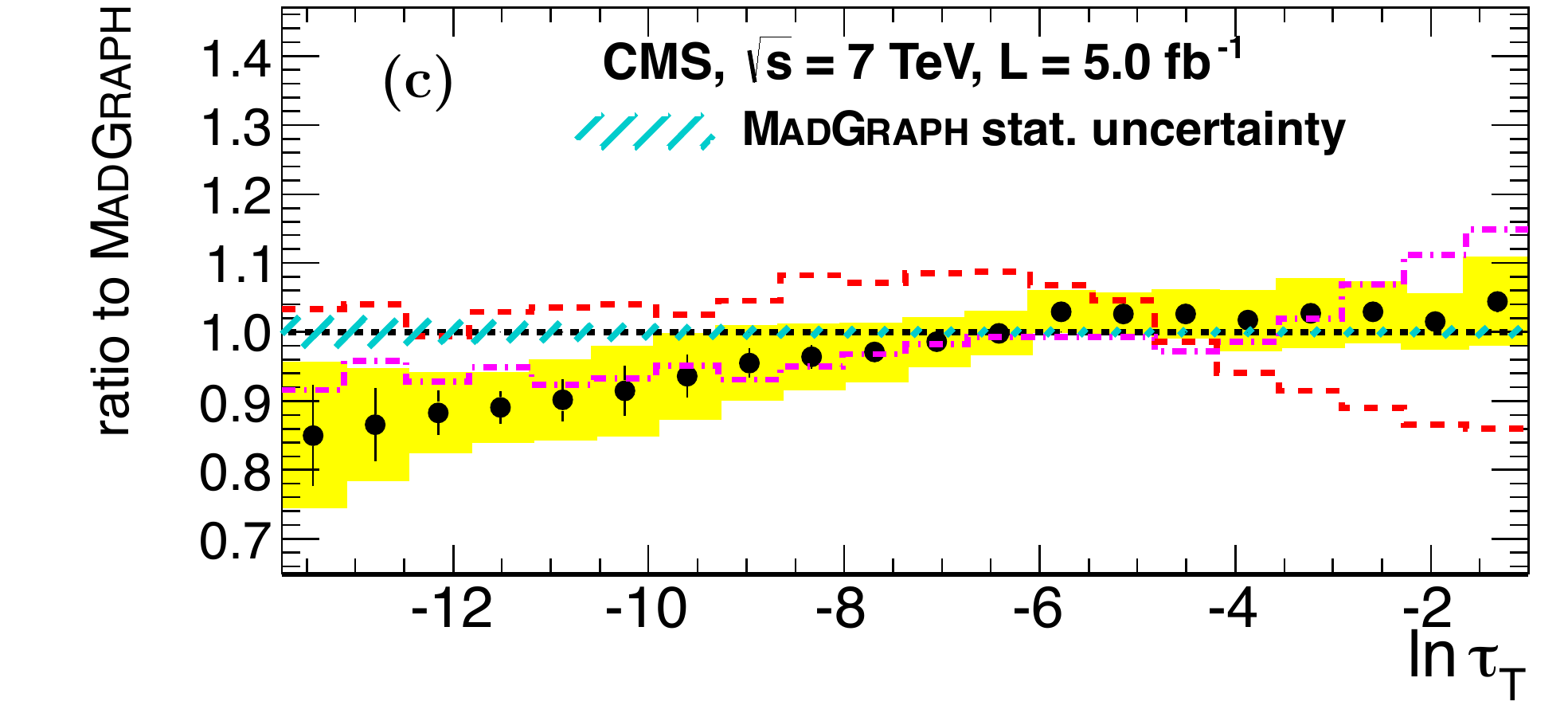}  {fig:thrust_results:Ratio}
    \setsubfloatTR{0.49\textwidth}{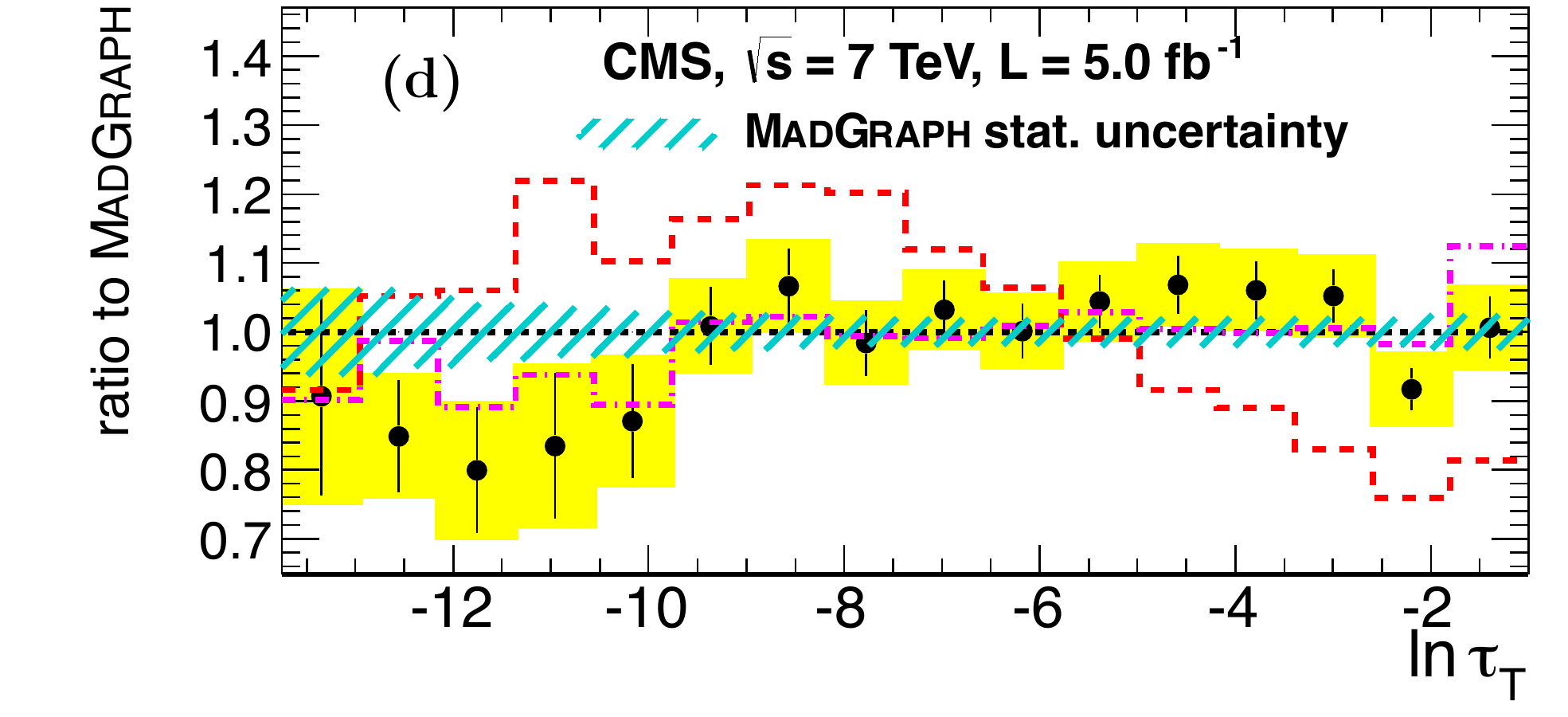}{fig:thrust_results:RatioBoost}
    \caption{
Normalized distributions in $\ln\tau_\mathrm{T}$ for (a) all $\ptZ$ and $\njets\geq1$  data, and (b) for $\ptZ>150\GeV$ and  $\njets\geq1$.
Plots in (c) and (d) show the ratios of the data and other \MC predictions, relative to \MADGRAPH,
as described in Fig.~\ref{fig:dphiZJ1jetmultbinned}.
}
    \label{fig:thrust_results}
  \end{center}
\end{figure*}

The corrected differential cross sections (normalized to unity) are compared to the predictions of \MADGRAPH, \SHERPA, \POWHEG  \Z + 1-jet (NLO), and \PYTHIA generators.
The  differential cross sections in the $\Delta\phi$ and $\ln\tau_\mathrm{T}$ variables are divided by the total \zjets cross section for the range defined by the lepton and jet kinematic
selection criteria, i.e. $\pt>20\GeV$ and $|\eta|<2.4$ for leptons, and $\pt>50\GeV$ and $|\eta|<2.5$ for jets, and $\Delta R>0.4$ for jet--lepton separation.
The distributions in data and MC are therefore normalized to unity.
Figure~\ref{fig:dphiZJ1jetmultbinned} shows $\Delta\phi(\cPZ,j_1)$ as a function of jet multiplicity for inclusive \Z + $\njets$ production, with  $\njets \geq 1,\ 2,$ and $3$.
Figures~\ref{fig:azimuth3jetspectra1} and~\ref{fig:azimuth3jetspectra2} show the $\Delta\phi(\cPZ,j_{i})$ and $\Delta\phi(j_{i},j_{k})$ distributions,
where $i,k$ represent jet indices in order of decreasing \pt, for $\njets\geq 3$.
For the sake of comparison, the $\Delta\phi(\cPZ,j_1)$ results for $\njets \geq 3$ from Fig.~\ref{fig:dphiZJ1jetmultbinned} are also included in Fig.~\ref{fig:azimuth3jetspectra1}.
These distributions characterize essentially all the azimuthal correlations in the $\njets\geq 3$ inclusive jet-multiplicity bin.
Finally, the \zjets distributions in $\ln\tau_\mathrm{T}$ are presented in Fig.~\ref{fig:thrust_results}.
For both azimuthal and $\ln\tau_\mathrm{T}$ distributions, the sum of statistical and systematic uncertainties taken in quadrature are presented as solid yellow shaded bands.
The statistical uncertainty from the \MADGRAPH \MC is displayed as a cross-hatched band for each distribution.

Overall, the measured distributions in $\Delta\phi(\cPZ,j_1)$ agree within uncertainties with the predictions from \MADGRAPH.
The predictions from \SHERPA underestimate the measured distributions by about $10\%$ whereas \POWHEG predictions overestimate by about $10\%$.
The disagreements with \SHERPA and \POWHEG (as well as between the two models) become less pronounced at larger inclusive jet multiplicities (Fig.~\ref{fig:dphiZJ1jetmultbinned}).
For $\njets=1$, the \Z boson and the accompanying parton are completely correlated, and $\Delta\phi(\cPZ,j_1) \approx \pi$ (Fig.~\ref{fig:cartoon}a).
When $\Delta\phi(\cPZ,j_1)\ll\pi$, the presence of additional hard QCD radiation is implied.
Certain configurations of jets with $\Delta\phi(\cPZ,j_1)<\pi/2$ probe events where the \Z boson is in the same hemisphere as the leading jet, and the $\vec{p}_\mathrm{T}$ of
the \Z boson is therefore balanced by at least two (or more) subleading jets emitted in the opposite hemisphere (Fig.~\ref{fig:cartoon}b).
The importance of the multiparton LO+PS approach, as reflected in \MADGRAPH and \SHERPA,
can be seen when the data are compared to stand-alone \PYTHIA at $\Delta\phi(\cPZ,j_1) < 2.5$ and $\njets\geq 1$.
For higher jet multiplicities of $\njets\geq 2$ and $\geq 3$, the distribution in $\Delta\phi(\cPZ,j_1)$ becomes more isotropic, although a strong correlation remains at $\Delta\phi=\pi$.

Within uncertainties, good agreement is observed between the data and \MADGRAPH, \SHERPA, and \POWHEG event generators for $\njets\geq3$.
Stand-alone \PYTHIA is also consistent with the distributions in $\Delta\phi(\cPZ,j_3)$ and $\Delta\phi(j_2,j_3)$.
In \PYTHIA, these high-multiplicity configurations are generated exclusively from the PS contribution. The important role of
the PS approximation in modifying the kinematics predicted from fixed-order calculations
is emphasized in \POWHEG, where
its predictive power in a multijet environment ($\njets \geq 3$)
is evident in Figs.~\ref{fig:dphiZJ1jetmultbinned}--\ref{fig:azimuth3jetspectra2}.
While \POWHEG represents an NLO prediction only for the leading jet, and additional radiation is modeled exclusively using parton showers, good agreement is observed
for data with $\njets\geq3$.

For the region $\ptZ>150\GeV$, the $\Delta\phi(\cPZ,j_1)$ distributions become more isotropic as jet multiplicity increases.
In addition, and contrary to the result for all $\ptZ$, the angular distributions between the subleading jets  $\Delta\phi(j_{i},j_{k})$ also become isotropic
(Fig.~\ref{fig:azimuth3jetspectra2}b).
The improved performance of \PYTHIA in this region is consistent with the increased phase space available for parton emission.
A similar observation can be made for distributions in $\ln\tau_\mathrm{T}$, which are discussed below.
The level of agreement between \PYTHIA and data for distributions in $\ln\tau_\mathrm{T}$ improves for $\ptZ>150$\GeV (Fig.~\ref{fig:thrust_results}).

The corrected normalized differential distributions in $\ln\tau_\mathrm{T}$ are displayed in Fig.~\ref{fig:thrust_results}.
The distributions for large $\ptZ$ indicate an accumulation of events at values of $\ln\tau_\mathrm{T}\approx-2$,
as could be expected, because this region of phase space corresponds to contributions from events with a large spherical component, corresponding to production of two or more jets.
Among the four examined models, \POWHEG and \MADGRAPH are more consistent with the data, being within 10\% of the measured distributions, except at large negative values
of $\ln\tau_\mathrm{T}$, where ${\approx}15$--$20\%$ deviations are observed.
The level of agreement of \SHERPA with data corresponds to 10--15\% for most of the bins, while \PYTHIA shows discrepancies of $>20\%$.
\PYTHIA and \SHERPA also predict too small values of $\ln\tau_\mathrm{T}$, especially at values dominated by configurations in which the leading jet is produced back-to-back with the \Z boson.
This yields a larger proportion of back-to-back \Z + 1-jet events relative to data at small $\ln\tau_\mathrm{T}$, an effect that  can also be observed in the $\Delta\phi(\cPZ,j_1)$ distribution
of Fig.~\ref{fig:dphiZJ1jetmultbinned}.

\section{Summary}
This Letter reports studies of angular correlations among the objects in \zjets final states.
The measurements are based on data corresponding to an integrated luminosity of \lumifinal, collected with the CMS detector at the LHC in proton-proton collisions at $\sqrt{s}=7\TeV$.

Azimuthal correlations among the \Z boson and the accompanying jets, $\Delta\phi(\cPZ,j_{i})$ and $\Delta\phi(j_{i},j_{k})$, are measured as functions of  inclusive jet multiplicity
($\njets\geq1$, $\geq 2$, and $\geq 3$).
In addition, the  transverse thrust event-shape variable $\ln\tau_\mathrm{T}$ is used to characterize the events.
Two regions of phase space are probed: (i) all events, independent of $\ptZ$, and (ii) the more highly boosted subset of events with $\ptZ>150\GeV$.
The systematic uncertainties are smaller than those arising from statistical sources, which dominate in the extreme regions of phase space.

The data are compared with predictions from \MADGRAPH, \SHERPA, \POWHEG \Z + 1-jet (at NLO), and stand-alone \PYTHIA \Z + 1-jet (at LO).
\PYTHIA corresponds to the simplest model, and is used to gauge the importance of additional corrections from LO and NLO ME formulations
that are interfaced with programs that evolve parton showers.
Stand-alone \PYTHIA provides an adequate description of event topologies when the phase space available for parton emission is large, \eg for the highly boosted selection on $\ptZ$.
The \MC models that combine multiparton QCD LO ME interfaced to parton shower evolution tend to agree with the data.
The  \Z + 1-jet ME calculation (at NLO) provided by \POWHEG shows agreement with data at large jet multiplicity in the entire phase space
probed in this study, despite the fact that, beyond the subleading jet, additional radiation comes exclusively from parton showers.

The measurements presented in this study provide a detailed description of the topological structure of \zjets production that is complementary to existing measurements of rates and
associated jet multiplicities.
As theoretical understanding evolves, these results can be used as additional probes of the validity of QCD predictions, while also providing confidence in the
current MC models as useful tools for the description of SM processes and their application for determining background in searches for new phenomena.

\section*{Acknowledgments}
We congratulate our colleagues in the CERN accelerator departments for the excellent performance of the LHC and thank the technical and administrative staffs at CERN and at other CMS institutes for their contributions to the success of the CMS effort. In addition, we gratefully acknowledge the computing centers and personnel of the Worldwide LHC Computing Grid for delivering so effectively the computing infrastructure essential to our analyses. Finally, we acknowledge the enduring support for the construction and operation of the LHC and the CMS detector provided by the following funding agencies: BMWF and FWF (Austria); FNRS and FWO (Belgium); CNPq, CAPES, FAPERJ, and FAPESP (Brazil); MEYS (Bulgaria); CERN; CAS, MoST, and NSFC (China); COLCIENCIAS (Colombia); MSES (Croatia); RPF (Cyprus); MoER, SF0690030s09 and ERDF (Estonia); Academy of Finland, MEC, and HIP (Finland); CEA and CNRS/IN2P3 (France); BMBF, DFG, and HGF (Germany); GSRT (Greece); OTKA and NKTH (Hungary); DAE and DST (India); IPM (Iran); SFI (Ireland); INFN (Italy); NRF and WCU (Republic of Korea); LAS (Lithuania); CINVESTAV, CONACYT, SEP, and UASLP-FAI (Mexico); MSI (New Zealand); PAEC (Pakistan); MSHE and NSC (Poland); FCT (Portugal); JINR (Armenia, Belarus, Georgia, Ukraine, Uzbekistan); MON, RosAtom, RAS and RFBR (Russia); MSTD (Serbia); SEIDI and CPAN (Spain); Swiss Funding Agencies (Switzerland); NSC (Taipei); ThEPCenter, IPST and NSTDA (Thailand); TUBITAK and TAEK (Turkey); NASU (Ukraine); STFC (United Kingdom); DOE and NSF (USA).

\bibliography{auto_generated}   % will be created by the tdr script.

\cleardoublepage \appendix\section{The CMS Collaboration \label{app:collab}}\begin{sloppypar}\hyphenpenalty=5000\widowpenalty=500\clubpenalty=5000\input{EWK-11-021-authorlist.tex}\end{sloppypar}
\end{document}

%% file: EWK-11-021-authorlist.tex
\textbf{Yerevan Physics Institute,  Yerevan,  Armenia}\\*[0pt]
S.~Chatrchyan, V.~Khachatryan, A.M.~Sirunyan, A.~Tumasyan
\vskip\cmsinstskip
\textbf{Institut f\"{u}r Hochenergiephysik der OeAW,  Wien,  Austria}\\*[0pt]
W.~Adam, E.~Aguilo, T.~Bergauer, M.~Dragicevic, J.~Er\"{o}, C.~Fabjan\cmsAuthorMark{1}, M.~Friedl, R.~Fr\"{u}hwirth\cmsAuthorMark{1}, V.M.~Ghete, N.~H\"{o}rmann, J.~Hrubec, M.~Jeitler\cmsAuthorMark{1}, W.~Kiesenhofer, V.~Kn\"{u}nz, M.~Krammer\cmsAuthorMark{1}, I.~Kr\"{a}tschmer, D.~Liko, I.~Mikulec, M.~Pernicka$^{\textrm{\dag}}$, D.~Rabady\cmsAuthorMark{2}, B.~Rahbaran, C.~Rohringer, H.~Rohringer, R.~Sch\"{o}fbeck, J.~Strauss, A.~Taurok, W.~Waltenberger, C.-E.~Wulz\cmsAuthorMark{1}
% ...
\vskip\cmsinstskip
\dag:~Deceased\\
1:~~Also at Vienna University of Technology, Vienna, Austria\\
2:~~Also at CERN, European Organization for Nuclear Research, Geneva, Switzerland\\
% ...


%% file: EWK-11-021_temp.bbl
\providecommand{\href}[2]{#2}\begingroup\raggedright\begin{thebibliography}{10}%
\makeatletter
\providecommand{\hrefCMSnoop }[0]{\@secondoftwo}%
\makeatother
\providecommand{\doi}{\texttt{doi:}\begingroup \urlstyle{tt}\Url}

\bibitem{Ita:2011wn}
H.~Ita\hrefCMSnoop {} { {et~al.}, ``Precise predictions for {Z}-boson + 4 jet
  production at hadron colliders'',} \textit{ Phys. Rev. D} \textbf{ 85} (2012)
  031501,
  \href{http://dx.doi.org/10.1103/PhysRevD.85.031501}{\doi{10.1103/PhysRevD.85.031501}}.

\bibitem{powheg:2004}
\hrefCMSnoop {} {P.~Nason, ``A new method for combining {NLO QCD} with shower
  {Monte Carlo} algorithms'',} \textit{ JHEP} \textbf{ 11} (2004) 040,
  \href{http://dx.doi.org/10.1088/1126-6708/2004/11/040}{\doi{10.1088/1126-6708/2004/11/040}}.

\bibitem{powheg:2007}
\hrefCMSnoop {} {S.~Frixione, P.~Nason, and C.~Oleari, ``Matching {NLO QCD}
  computations with partonshower simulations: the {POWHEG} method'',} \textit{
  JHEP} \textbf{ 11} (2007) 070,
  \href{http://dx.doi.org/10.1088/1126-6708/2007/11/070}{\doi{10.1088/1126-6708/2007/11/070}}.

\bibitem{Alioli:2010xd}
S.~Alioli\hrefCMSnoop {} { {et~al.}, ``A general framework for implementing
  {NLO} calculations in shower {Monte Carlo} programs: the {POWHEG BOX}'',}
  \textit{ JHEP} \textbf{ 06} (2010) 043,
  \href{http://dx.doi.org/10.1007/JHEP06(2010)043}{\doi{10.1007/JHEP06(2010)043}}.

\bibitem{Alioli:2010qp}
S.~Alioli\hrefCMSnoop {} { {et~al.}, ``Vector boson plus one jet production in
  {POWHEG}'',} \textit{ JHEP} \textbf{ 01} (2011) 095,
  \href{http://dx.doi.org/10.1007/JHEP01(2011)095}{\doi{10.1007/JHEP01(2011)095}}.

\bibitem{MCatNLO}
\hrefCMSnoop {} {S.~Frixione and B.~R. Webber, ``Matching {NLO QCD}
  computations and parton shower simulations'',} \textit{ JHEP} \textbf{ 06}
  (2002) 029,
  \href{http://dx.doi.org/10.1088/1126-6708/2002/06/029}{\doi{10.1088/1126-6708/2002/06/029}}.

\bibitem{Mangano:2002ea}
M.~L. Mangano\hrefCMSnoop {} { {et~al.}, ``{ALPGEN}, a generator for hard
  multiparton processes in hadronic collisions'',} \textit{ JHEP} \textbf{ 07}
  (2003) 001,
  \href{http://dx.doi.org/10.1088/1126-6708/2003/07/001}{\doi{10.1088/1126-6708/2003/07/001}}.

\bibitem{Alwall:2011uj}
J.~Alwall\hrefCMSnoop {} { {et~al.}, ``{MadGraph} 5: going beyond'',} \textit{
  JHEP} \textbf{ 06} (2011) 128,
  \href{http://dx.doi.org/10.1007/JHEP06(2011)128}{\doi{10.1007/JHEP06(2011)128}}.

\bibitem{sherpa}
T.~Gleisberg\hrefCMSnoop {} { {et~al.}, ``Event generation with {SHERPA}
  1.1'',} \textit{ JHEP} \textbf{ 02} (2009) 007,
  \href{http://dx.doi.org/10.1088/1126-6708/2009/02/007}{\doi{10.1088/1126-6708/2009/02/007}}.

\bibitem{d0vjets}
\hrefCMSnoop {} {{ D0} Collaboration, ``Measurement of {$Z /
  \gamma^{*}$+jet+$X$} angular distributions in $p\bar{p}$ collisions at
  $\sqrt{s}$ = 1.96 {TeV}'',} \textit{ Phys. Lett. B} \textbf{ 682} (2010) 370,
  \href{http://dx.doi.org/10.1016/j.physletb.2009.11.012}{\doi{10.1016/j.physletb.2009.11.012}}.

\bibitem{atlasvjets}
\hrefCMSnoop {} {{ ATLAS} Collaboration, ``Measurement of the production cross
  section for {$Z/\gamma^{*}$} in association with jets in $pp$ collisions at
  $\sqrt{s}=7$ $\mathrm{TeV}$ with the {ATLAS} detector'',} \textit{ Phys. Rev.
  D} \textbf{ 85} (2012) 032009,
  \href{http://dx.doi.org/10.1103/PhysRevD.85.032009}{\doi{10.1103/PhysRevD.85.032009}}.

\bibitem{H1evshp}
\hrefCMSnoop {} {{ H1} Collaboration, ``Measurement of event shape variables in
  deep-inelastic scattering at {HERA}'',} \textit{ Eur. Phys. J. C} \textbf{
  46} (2006) 343,
  \href{http://dx.doi.org/10.1140/epjc/s2006-02493-x}{\doi{10.1140/epjc/s2006-02493-x}}.

\bibitem{Zeusevshp}
\hrefCMSnoop {} {{ ZEUS} Collaboration, ``Event shapes in deep inelastic
  scattering at HERA'',} \textit{ Nucl. Phys. B} \textbf{ 767} (2007) 1,
  \href{http://dx.doi.org/10.1016/j.nuclphysb.2006.05.016}{\doi{10.1016/j.nuclphysb.2006.05.016}}.

\bibitem{DelphiTuningMCevshp}
\hrefCMSnoop {} {{ DELPHI} Collaboration, ``Tuning and test of fragmentation
  models based on identified particles and precision event shape data'',}
  \textit{ Z. Phys. C} \textbf{ 73} (1996) 11,
  \href{http://dx.doi.org/10.1007/s002880050295}{\doi{10.1007/s002880050295}}.

\bibitem{dissertorievshp}
G.~Dissertori\hrefCMSnoop {} { {et~al.}, ``Determination of the strong coupling
  constant using matched {NNLO+NLLA} predictions for hadronic event shapes in
  {\EE} annihilations'',} \textit{ JHEP} \textbf{ 08} (2009) 036,
  \href{http://dx.doi.org/10.1088/1126-6708/2009/08/036}{\doi{10.1088/1126-6708/2009/08/036}}.

\bibitem{AlephQCDoverview}
\hrefCMSnoop {} {{ ALEPH} Collaboration, ``Studies of quantum chromodynamics
  with the {ALEPH} detector'',} \textit{ Phys. Rept.} \textbf{ 294} (1998) 1,
  \href{http://dx.doi.org/10.1016/S0370-1573(97)00045-8}{\doi{10.1016/S0370-1573(97)00045-8}}.

\bibitem{evshpSalam}
\hrefCMSnoop {} {A.~Banfi, G.~P. Salam, and G.~Zanderighi, ``Resummed event
  shapes at hadron--hadron colliders'',} \textit{ JHEP} \textbf{ 08} (2004)
  062,
  \href{http://dx.doi.org/10.1088/1126-6708/2004/08/062}{\doi{10.1088/1126-6708/2004/08/062}}.

\bibitem{evshpPaper}
\hrefCMSnoop {} {{ CMS} Collaboration, ``First measurement of hadronic event
  shapes in pp collisions at $\sqrt{s} = 7\,\mathrm{TeV}$'',} \textit{ Phys.
  Lett. B} \textbf{ 699} (2011) 48,
  \href{http://dx.doi.org/10.1016/j.physletb.2011.03.060}{\doi{10.1016/j.physletb.2011.03.060}}.

\bibitem{Chatrchyan:2008zzk}
\hrefCMSnoop {} {{ CMS} Collaboration, ``The {CMS} experiment at the {CERN}
  {LHC}'',} \textit{ JINST} \textbf{ 3} (2008) S08004,
  \href{http://dx.doi.org/10.1088/1748-0221/3/08/S08004}{\doi{10.1088/1748-0221/3/08/S08004}}.

\bibitem{pythia6}
\hrefCMSnoop {} {T.~Sj{\"o}strand, S.~Mrenna, and P.~Skands, ``{PYTHIA} 6.4
  physics and manual'',} \textit{ JHEP} \textbf{ 05} (2006) 026,
  \href{http://dx.doi.org/10.1088/1126-6708/2006/05/026}{\doi{10.1088/1126-6708/2006/05/026}}.

\bibitem{Chatrchyan:2011id}
\hrefCMSnoop {} {{ CMS} Collaboration, ``{Measurement of the underlying event
  activity at the LHC with $\sqrt{s} = 7$ TeV and comparison with $\sqrt{s} =
  0.9$ TeV}'',} \textit{ JHEP} \textbf{ 09} (2011) 109,
  \href{http://dx.doi.org/10.1007/JHEP09(2011)109}{\doi{10.1007/JHEP09(2011)109}}.

\bibitem{Pumplin:2002vw}
J.~Pumplin\hrefCMSnoop {} { {et~al.}, ``New generation of parton distributions
  with uncertainties from global {QCD} analysis'',} \textit{ JHEP} \textbf{ 07}
  (2002) 012,
  \href{http://dx.doi.org/10.1088/1126-6708/2002/07/012}{\doi{10.1088/1126-6708/2002/07/012}}.

\bibitem{Lai:2010vv}
H.~L. Lai\hrefCMSnoop {} { {et~al.}, ``New parton distributions for collider
  physics'',} \textit{ Phys. Rev. D} \textbf{ 82} (2010) 074024,
  \href{http://dx.doi.org/10.1103/PhysRevD.82.074024}{\doi{10.1103/PhysRevD.82.074024}}.

\bibitem{nnlo:Z}
\hrefCMSnoop {} {K.~Melnikov and F.~Petriello, ``Electroweak gauge boson
  production at hadron colliders through ${O}(\alpha_{S}^{2})$'',} \textit{
  Phys. Rev. D} \textbf{ 74} (2006) 114017,
  \href{http://dx.doi.org/10.1103/PhysRevD.74.114017}{\doi{10.1103/PhysRevD.74.114017}}.

\bibitem{ttbarxsec}
\hrefCMSnoop {} {N.~Kidonakis, ``Higher-order corrections to top-antitop pair
  and single top quark production'',} (2009).
  \href{http://www.arXiv.org/abs/0909.0037}{\texttt{ arXiv:0909.0037}}.

\bibitem{Campbell:2010ff}
\hrefCMSnoop {} {J.~M. Campbell and R.~K. Ellis, ``{MCFM for the Tevatron and
  the LHC}'',} \textit{ Nucl. Phys. Proc. Suppl.} \textbf{ 205-206} (2010) 10,
  \href{http://dx.doi.org/10.1016/j.nuclphysbps.2010.08.011}{\doi{10.1016/j.nuclphysbps.2010.08.011}},
\href{http://www.arXiv.org/abs/1007.3492}{\texttt{ arXiv:1007.3492}}.
%%CITATION = ARXIV:1007.3492;%%.

\bibitem{geant4}
\hrefCMSnoop {} {{ GEANT4} Collaboration, ``{GEANT}4---a simulation toolkit'',}
  \textit{ Nucl. Instrum. Meth. A} \textbf{ 506} (2003) 250,
  \href{http://dx.doi.org/10.1016/S0168-9002(03)01368-8}{\doi{10.1016/S0168-9002(03)01368-8}}.

\bibitem{Chatrchyan:2012xi}
\hrefCMSnoop {} {{ CMS} Collaboration, ``{Performance of CMS muon
  reconstruction in pp collision events at $\sqrt{s}$ = 7 TeV}'',} \textit{
  JINST} \textbf{ 7} (2012) P10002,
  \href{http://dx.doi.org/10.1088/1748-0221/7/10/P10002}{\doi{10.1088/1748-0221/7/10/P10002}}.

\bibitem{CMS-PAS-EGM-10-004}
\href {https://cdsweb.cern.ch/record/1299116} {{ CMS} Collaboration, ``Electron
  Reconstruction and Identification at $\sqrt{s} = 7$ TeV'',} CMS Physics
  Analysis Summary CMS-PAS-EGM-10-004, (2010).

\bibitem{Cacciari:2007fd}
\hrefCMSnoop {} {M.~Cacciari and G.~P. Salam, ``Pileup subtraction using jet
  areas'',} \textit{ Phys. Lett. B} \textbf{ 659} (2008) 119,
  \href{http://dx.doi.org/10.1016/j.physletb.2007.09.077}{\doi{10.1016/j.physletb.2007.09.077}}.

\bibitem{inclusiveWZ}
\hrefCMSnoop {} {{ CMS} Collaboration, ``Measurement of the inclusive {W} and
  {Z} production cross sections in pp collisions at $\sqrt{s}$ = 7 {TeV} with
  the {CMS} experiment'',} \textit{ JHEP} \textbf{ 10} (2011) 132,
  \href{http://dx.doi.org/10.1007/JHEP10(2011)132}{\doi{10.1007/JHEP10(2011)132}}.

\bibitem{CMS-PAS-PFT-09-001}
\href {http://cdsweb.cern.ch/record/1194487} {{ CMS} Collaboration,
  ``Particle-Flow Event Reconstruction in {CMS} and Performance for Jets, Taus,
  and {\MET}'',} CMS Physics Analysis Summary CMS-PAS-PFT-09-001, (2009).

\bibitem{CMS-PAS-PFT-10-002}
\href {http://cdsweb.cern.ch/record/1247373} {{ CMS} Collaboration,
  ``Commissioning of the Particle-Flow Event Reconstruction with the first
  {LHC} collisions recorded in the {CMS} detector'',} CMS Physics Analysis
  Summary CMS-PAS-PFT-10-002, (2010).

\bibitem{Cacciari:2008gp}
\hrefCMSnoop {} {M.~Cacciari, G.~P. Salam, and G.~Soyez, ``The anti-$k_t$ jet
  clustering algorithm'',} \textit{ JHEP} \textbf{ 04} (2008) 063,
  \href{http://dx.doi.org/10.1088/1126-6708/2008/04/063}{\doi{10.1088/1126-6708/2008/04/063}}.

\bibitem{fastjet1}
\hrefCMSnoop {} {M.~Cacciari and G.~P. Salam, ``Dispelling the ${N}^{3}$ myth
  for the $k_t$ jet-finder'',} \textit{ Phys. Lett. B} \textbf{ 641} (2006) 57,
  \href{http://dx.doi.org/10.1016/j.physletb.2006.08.037}{\doi{10.1016/j.physletb.2006.08.037}}.

\bibitem{fastjet2}
\hrefCMSnoop {} {M.~Cacciari, G.~P. Salam, and G.~Soyez, ``{FastJet} user
  manual (for version 3.0.2)'',} \textit{ Eur. Phys. J. C} \textbf{ 72} (2012)
  1896,
  \href{http://dx.doi.org/10.1140/epjc/s10052-012-1896-2}{\doi{10.1140/epjc/s10052-012-1896-2}}.

\bibitem{JES2011}
\hrefCMSnoop {} {{ CMS} Collaboration, ``Determination of jet energy
  calibration and transverse momentum resolution in {CMS}'',} \textit{ JINST}
  \textbf{ 6} (2011) P11002,
  \href{http://dx.doi.org/10.1088/1748-0221/6/11/P11002}{\doi{10.1088/1748-0221/6/11/P11002}}.

\bibitem{Buttar:2008jx}
\hrefCMSnoop {} {C.~Buttar {et~al.}, ``{Standard Model Handles and Candles
  Working Group: Tools and Jets Summary Report}'',} (2008).
\href{http://www.arXiv.org/abs/0803.0678}{\texttt{ arXiv:0803.0678}}.
%%CITATION = ARXIV:0803.0678;%%.

\bibitem{CMS-PAS-SMP-12-008}
\href {https://cdsweb.cern.ch/record/1434360} {{ CMS} Collaboration, ``Absolute
  Calibration of the Luminosity Measurement at CMS: Winter 2012 Update'',} CMS
  Physics Analysis Summary CMS-PAS-SMP-12-008, (2012).

\bibitem{BLUE}
\hrefCMSnoop {} {L.~Lyons, D.~Gibaut, and P.~Clifford, ``{How to combine
  correlated estimates of a single physical quantity}'',} \textit{ Nucl.
  Instrum. Meth. A} \textbf{ 270} (1988) 110,
  \href{http://dx.doi.org/10.1016/0168-9002(88)90018-6}{\doi{10.1016/0168-9002(88)90018-6}}.

\bibitem{SVD}
\hrefCMSnoop {} {A.~H{\"o}cker and V.~Kartvelishvili, ``{SVD} approach to data
  unfolding'',} \textit{ Nucl. Instrum. Meth. A} \textbf{ 372} (1996) 469,
  \href{http://dx.doi.org/10.1016/0168-9002(95)01478-0}{\doi{10.1016/0168-9002(95)01478-0}}.

\bibitem{RooUnfold}
\hrefCMSnoop {} {T.~Adye, ``Unfolding algorithms and tests using
  {RooUnfold}'',} (2011). \href{http://www.arXiv.org/abs/1105.1160}{\texttt{
  arXiv:1105.1160}}.

\end{thebibliography}\endgroup
